\documentclass[twocolumn, numberedappendix, twocolappendix, apj]{openjournal}
\usepackage{newtxtext,newtxmath}
\usepackage{amsmath}
\usepackage{amssymb}
\usepackage{graphicx}
\usepackage{xspace}
\usepackage[table]{xcolor}
\usepackage[colorlinks=true, allcolors=blue]{hyperref}
\usepackage{fontawesome5}
\usepackage{multirow}
\usepackage{listings}
\usepackage{xcolor}

\newcommand{\sxt}{7$\times$2-pt\xspace}
\newcommand{\LB}{\texttt{LimberJack.jl}\xspace}
\newcommand{\jl}{\texttt{Julia}\xspace}
\newcommand{\lcdm}{$\Lambda$CDM\xspace}

\definecolor{codegreen}{rgb}{0,0.6,0}
\definecolor{codegray}{rgb}{0.5,0.5,0.5}
\definecolor{codepurple}{rgb}{0.58,0,0.82}
\definecolor{backcolour}{rgb}{0.95,0.95,0.92}
\lstdefinestyle{mystyle}{
    backgroundcolor=\color{backcolour},   
    commentstyle=\color{codegreen},
    keywordstyle=\color{magenta},
    numberstyle=\tiny\color{codegray},
    stringstyle=\color{codepurple},
    basicstyle=\ttfamily\footnotesize,
    breakatwhitespace=false,         
    breaklines=true,                 
    captionpos=b,                    
    keepspaces=true,                 
    numbers=left,                    
    numbersep=5pt,                  
    showspaces=false,                
    showstringspaces=false,
    showtabs=false,                  
    tabsize=2
}

\lstdefinelanguage{Julia}%
  {morekeywords={abstract,break,case,catch,const,continue,do,else,elseif,%
      end,export,false,for,function,immutable,import,importall,if,in,%
      macro,module,otherwise,quote,return,switch,true,try,type,typealias,%
      using,while},%
   sensitive=true,%
   morecomment=[l]\#,%
   morecomment=[n]{\#=}{=\#},%
   morestring=[s]{"}{"},%
   morestring=[m]{'}{'},%
}[keywords,comments,strings]%

\begin{document}

\title{LimberJack.jl: auto-differentiable methods for angular power spectra analyses}
\author{
J. Ruiz-Zapatero$^{1,*}$,
D. Alonso$^{1}$,
C. Garc\'ia-Garc\'ia$^1$,
A. Nicola$^2$,
A. Mootoovaloo$^1$,
J. M. Sullivan$^{3,4}$,
M. Bonici$^5$,
P. G. Ferreira$^1$}
\email{$^*$jaime.ruiz-zapatero@physics.ox.ac.uk}

\affiliation{$^1$Astrophysics, University of Oxford, DWB, Keble Road, Oxford OX1 3RH, UK}
\affiliation{$^2$Argelander Institut für Astronomie, Universität Bonn, Auf dem Hügel 71, 53121 Bonn, Germany}
\affiliation{$^3$Department of Astronomy, University of California, Berkeley, CA 94720, USA}
\affiliation{$^4$Berkeley Center for Cosmological Physics, University of California, Berkeley, CA 94720,USA} 
\affiliation{$^5$INAF-IASF, Milano, Italy}
\date{\today}

\begin{abstract}
We present \LB, a fully auto-differentiable code for cosmological analyses of 2 point auto- and cross-correlation measurements from galaxy clustering, CMB lensing and weak lensing data written in \jl. Using \jl's auto-differentiation ecosystem, \LB can obtain gradients for its outputs an order of magnitude faster than traditional finite difference methods. This makes \LB greatly synergistic with gradient-based sampling methods, such as Hamiltonian Monte Carlo, capable of efficiently exploring parameter spaces with hundreds of dimensions. We first prove \LB's reliability by reanalysing the DES Y1 3$\times$2-point data. We then showcase its capabilities by using a O(100) parameters Gaussian Process to reconstruct the cosmic growth from a combination of DES Y1 galaxy clustering and weak lensing data, eBOSS QSO's, CMB lensing and redshift-space distortions. Our Gaussian process reconstruction of the growth factor is statistically consistent with the \lcdm Planck 2018 prediction at all redshifts. Moreover, we show that the addition of RSD data is extremely beneficial to this type of analysis, reducing the uncertainty in the reconstructed growth factor by $20\%$ on average across redshift.  \LB is a fully open-source project available on \jl's general repository of packages and \href{https://github.com/jaimerzp/LimberJack.jl}{GitHub} \textcolor{blue}{\faGithub}.  
\end{abstract}
\keywords{Large scale structure, cosmology, statistical inference}

\maketitle


\section{Introduction} \label{Sect: introduction}

Cosmology is currently experiencing an unprecedented increase in the quantity and quality of data. The Dark Energy Spectroscopic Instrument (DESI) \citep{DESI} has already recorded more galaxy spectra in its first two years of operations than previously done in the whole history of humanity. Moreover, next-generation surveys such as Legacy Survey of Space and Time (LSST) \citep{LSST}, Euclid \citep{Euclid}, the Nancy Grace Roman space telescope \citep{NGR_telescope}, the Simons Observatory \citep{Simons_observatory}, CMB-HD \citep{CMB_HD} or CMB-S4 \citep{CMB_S4} promise to further accelerate this trend in the next decade. 

In order to match the quality of the data, physicists are starting to incorporate into their theoretical predictions more of the physical, observational and instrumental effects which, until now, could be overlooked. In practice this translates into a dramatic increase in the number of parameters that future analyses will have to consider. This combination of large data sets with complex models will (and in many cases already does) overwhelm the inference methods we currently use to constrain the values of these parameters. 

In a Bayesian framework, the statistical distribution of a set of parameters, $\boldsymbol{\theta}$, given some data, $\boldsymbol{d}$, is given by Bayes theorem: 
\begin{equation} \label{eq. Bayes}
    \mathcal{P}(\boldsymbol{\theta}|\boldsymbol{d}) \propto \mathcal{L}(\boldsymbol{d}|\boldsymbol{\theta}) \Pi(\boldsymbol{\theta}) \, ,
\end{equation}
where $P(\boldsymbol{\theta}|\boldsymbol{d})$ is the posterior distribution of the parameters given the data, $\mathcal{L}(\boldsymbol{d}|\boldsymbol{\theta})$ is the likelihood of the data for a set of parameters and $\Pi(\boldsymbol{\theta})$ are the prior beliefs on the distribution of the parameters. 
Current cosmological analyses explore their parameter spaces by mapping out $P(\boldsymbol{\theta}|\boldsymbol{d})$ according to some stochastic process by which the direction of exploration is chosen. Despite the success of this methodology, relying on stochastic methods becomes inefficient at high dimensions as the chances of randomly finding the parameters that describe the observed data well decreases with the volume of the parameter space being probed. This effect is known as the "curse of dimensionality". 

One of the most effective ways of overcoming the curse of dimensionality is using the gradient of the posterior; $\nabla{P(\boldsymbol{\theta}|\boldsymbol{d})} \propto \nabla{(\mathcal{L}(\boldsymbol{d}|\boldsymbol{\theta}) \, \Pi(\boldsymbol{\theta}))}$, to guide exploration towards regions of interest in parameter space. Algorithms that use the gradient of the likelihood to explore parameter space are known as gradient-based samplers \citep{HMC, NUTS, PathFinder, MCHMC}. Unfortunately, traditional methods of finding numerical gradients, such as the finite-difference method, scale poorly with the number of dimensions. Moreover, they can be prone to numerical instabilities. This can render the use of gradient information counter-productive.

Thankfully, in the last decades a series of algorithms known as auto-differentiation (AD) \citep{evaluating_derivatives, AD_2, AD} have grown in popularity. Given a generic computer program that maps a series of inputs to a series of outputs (i.e. a function inside a computer) AD is a family of algorithms designed to produce a symbolic representation of such computer program such that the chain rule can be systematically applied to produce a second program for the gradient of the original function. Unlike finite differences, AD is not subject to truncation errors and its computational cost scales much more favourably with the dimensionality of the original function \citep{evaluating_derivatives}. Thus, the goal of this paper is to provide an AD cosmological inference framework.

We present \LB, an auto-differentiable angular power spectra analysis code fully written in \jl. \LB is designed after the Core Cosmology Library \citep[CCL][]{ccl} developed by the LSST Dark Energy Science Collaboration, aiming to fulfil the similar scientific goals. \LB allows the user to easily compute $\Lambda$CDM model predictions for the angular power spectra of weak lensing, CMB lensing and clustering surveys using the Limber approximation \citep{1953ApJ...117..134L, Extended_Limber}. Most importantly, \LB can also provide the user with accurate gradients of these predictions in a computationally efficient way, due to its compatibility with \jl's AD libraries \texttt{ForwardDiff.jl} and \texttt{ReverseDiff.jl}. While \LB is currently more limited than \texttt{CCL}, its modular design means that it can easily be extended by the community. This will allow the community to add new features that will be necessary for the analysis of future data which are currently not present in \LB; such as more precise prescriptions of the non-linear corrections to the matter power spectrum or going beyond the Limber approximation \citep{NK5}.

A number of auto-differentiable cosmological codes have been presented in the literature. The most notable precedents are \texttt{JAX-COSMO} \citep{JAX-COSMO} and \texttt{CosmoPower-JAX} \citep{CosmoPower-JAX}. \texttt{JAX-COSMO}'s functionalities and goals are very similar to those of \LB both basing their design on \texttt{CCL}. \texttt{CosmoPower-JAX} is a neural-network emulating framework for the matter power spectrum originally developed in \texttt{TensorFlow} \citep{CosmoPower} and later ported to \texttt{JAX} to be paired with \texttt{JAX-COSMO}. The main difference between  \texttt{JAX-COSMO}, \texttt{CosmoPower-JAX} and \LB is the AD ecosystem they make use of. Both \texttt{JAX-COSMO} and \texttt{CosmoPower-JAX} are written in \texttt{JAX}, a scripting programming language which interfaces with a lower-level language, \texttt{XLA}, a just-in-time (JIT) compiled language with AD capabilities. \texttt{JAX}'s main strengths are its powerful parallelisation schemes on both CPU's and GPU's, its performant AD methods and the fact that it shares API with the ubiquitous \texttt{Python} library \texttt{NumPy}. On the other hand, \LB is fully written in \jl, a general-purpose, JIT-compiled, programming language with native AD capabilities. \jl's main advantage is the lack of a lower-level programming language as is in the case of other  popular AD environments such as \texttt{TensorFlow} interfacing with \texttt{C++} and \texttt{JAX} with \texttt{XLA}. This transparency makes \jl an excellent language to develop complex libraries with customised methods. A recent practical example where this feature has played a key role is the development of the first AD Boltzmann solver, \texttt{Bolt.jl} \cite{Bolt}, a challenging task for \texttt{JAX} and \texttt{TensorFlow} but which, nonetheless, was possible in \jl.

The structure of this paper is as follows:  in Sect. \ref{Sect: Methods} we describe the theoretical predictions computed by \LB , how \LB computes these predictions in an auto-differentiable way and how AD can be used to speed up statistical inference. In Sect. \ref{Sect: Results}, we use \LB to reproduce the DES Y1 3x2-pt analysis and to perform a Gaussian process reconstruction of the growth factor across redshift. Finally, in Sect. \ref{Sect: Conclusions}, we summarise our work and interpret our results. In addition to this we also present the full derivatives of the most relevant theoretical predictions of \LB in App. \ref{App: Full Derivatives} and a tutorial on how to use \LB in App. \ref{App: Tutorial}.

\section{Methods} \label{Sect: Methods}
\subsection{Angular Power Spectra} \label{Subsect: Angular Power Spectra}
The main goal of \LB is to compute the \lcdm model predictions for the angular power spectra of any two tracers using the Limber approximation as well as their gradients. In the Limber approximation \citep{1953ApJ...117..134L, Extended_Limber}, the angular power spectrum, $C^{UV}_\ell$, between two projected fields $U$ and $V$ is related to the three-dimensional power spectrum, $P_{UV}(k,z)$ of their associated quantities $U$ and $V$ via:
\begin{equation}\label{eq:limber}
C^{UV}_\ell=\int\frac{d\chi}{\chi^2}q_U(\chi)q_V(\chi)\,P_{UV}\left(k=\frac{\ell+1/2}{\chi},z(\chi)\right) \, , 
\end{equation}
where $q_U$ and $q_V$ are the radial kernels of the $U$ and $V$ fields respectively.

Currently \texttt{LimberJack} can compute the auto- and cross-correlation of three different fields: cosmic shear  (i.e. the distortion in the shape of galaxies caused by weak gravitational lensing), CMB lensing convergence (i.e. the apparent magnification or reduction of CMB anisotropies due to gravitational lensing by the matter distribution along the line of sight) and galaxy clustering.

The radial kernels cosmic shear  ($q_\gamma$), CMB lensing convergence ($q_\kappa$) and galaxy clustering ($q_g$) are given by:
\begin{align}  \label{eq:kernels} 
  &q_\gamma(\chi) \equiv  \frac{3}{2}H_0^2\Omega_{\rm{m}}\frac{\chi}{a(\chi)}\int_{z(\chi)}^\infty dz' p(z')\frac{\chi(z')-\chi}{\chi(z')},  \\
  &q_\kappa(\chi) \equiv  \frac{3}{2}H_0^2\Omega_{\rm{m}}\frac{\chi}{a(\chi)}\frac{\chi^*-\chi}{\chi^*},\\
  &q_g(\chi) \equiv  \frac{H(z)}{c}p(z)
\end{align}
where $c$ is the speed of light, $H(z)$ is the Hubble expansion rate,  $H_0\equiv H(z=0)$, $\chi$ is the comoving distance, $a(\chi)$ is the expansion factor, $\Omega_{\rm{m}}$ is the matter density parameter today, $b_g$ is the galaxy-matter linear bias parameter, $\chi^*$ is the comoving distance to the surface of last scattering and $p(z)$ is the redshift distribution of the galaxies. 

The estimation of the galaxy clustering, shear and convergence fields is subject to a variety of systematic biases that must be accounted for in their modelling. In galaxy clustering analyses the level of detail to which the relation between the observed galaxy overdensity and the predicted matter overdensity  is modelled limits the range of scales that can be analysed. We consider a single parameter linear bias model, $b_{\rm{g}}$, that multiplies the clustering kernel. This restricts our analysis of galaxy clustering data to relatively large scales. Cosmic shear analysis must account for possible biases in galaxy shape estimation. Similarly, we consider a single multiplicative bias parameter, $m$. In addition, galaxy shapes are not only correlated by cosmic shear but also by intrinsic alignments (IAs) in the orientation of galaxies due to local interactions. Within the so-called Non-Linear Alignment \citep[NLA, ][]{NLA} model this can be accounted for by adding an extra contribution to the final shear kernel given by:
\begin{equation}
  q_I(\chi)= A_{\rm{IA}}(z) \frac{H(z)}{c}p(z), 
\end{equation}
where $A_{\rm{IA}}(z)$ is:
\begin{equation} \label{Eq.IA_amplitude}
    A_{\rm IA}(z)=A_{{\rm IA},0}\left( \frac{1+z}{1+z_0} \right)^{\eta_\text{IA}}\frac{0.0139 \Omega_{\rm{m}}}{D(z)}\, , 
\end{equation}
where $A_{\text{IA},0}$ and $\eta_{\text{IA}}$ are two free parameters, $z_0$ is a redshift pivot (which we fix to $z_0=0.62$ as in \cite{DESY1, 1708.01538}), and $D(z)$ the linear growth factor.

When the shear and convergence kernels enter Eq. \ref{eq:limber} they must be multiplied by an $\ell$-dependent correction to account for the differences between angular and three-dimensional derivatives of the matter density field. These corrections are given by:
\begin{align} 
  &W_{\rm{\ell}} \equiv \sqrt{\frac{(\ell+2)!}{(\ell-2)!}}\frac{1}{(\ell+1/2)^2} \,  \label{eq:ell_correction_1}  \\ 
  &K_{\rm{\ell}} \equiv \frac{\ell(\ell+1)}{(\ell+1/2)^2}   \,  \label{eq:ell_correction_2} 
\end{align}
for the shear and convergence kernels respectively.

Since all the fields considered are ultimately projections of the 3-dimensional matter density field we only need to consider the matter auto-correlation power spectrum, $P(k,z) \equiv P_{mm}(k,z) $, when computing Eq. \ref{eq:limber}. Assuming that the evolution of the matter power spectrum is scale independent, $P(k,z)$ can be obtained by first computing the value of the linear matter power spectrum at $z=0$ at an array of scales. As we will discuss in Sect. \ref{Subsect: limeberJack} this can be done either using fitting formulas or emulating the solutions of Botlzmann codes. Then, the linear matter power spectrum can be evolved backwards in time as $P^{\rm{L}}(k,z) = D^2(z) P_0(k)$ where $D(z)$ is the linear growth factor found by solving the Jeans equation:
\begin{equation} \label{eq:jeans_D}
    \frac{d}{da}\left(a^2H\frac{dD}{d\log a}\right)=\frac{3}{2}\Omega_{\rm{m}}(a)\,a\,H\,D \, .
\end{equation}
However, in order to fit the small scales it is necessary to compute the non-linear evolution of the matter power spectrum, $P^{\rm{NL}}(k,z)$. We do so by applying  the {\tt Halofit} fitting function \cite{Halofit_1} with revisions from \cite{Halofit_2} to $P^{\rm{L}}(k,z)$. However, analyses of data from the latest surveys \citep[See][among others]{DES_baryons, DES_K1K, HSC_DR3} require the use of the more accurate methods such as \texttt{HMCode} \citep{HMCode_og} or \texttt{baccoemu} \citep{BACCO_emulator} as smaller scales are included.

\subsection{LimberJack.jl} \label{Subsect: limeberJack}

The key feature of \LB which sets it apart from similar, more extensively tested codes (such as $\texttt{CCL}$ \citet{ccl}) are its AD capabilities. AD methods can be classified into two groups, forwards and backwards. In order to understand the difference, let us consider a complicated function of an independent variable, $f(x)$, one can represent the function as a composition of simpler functions, $f(x) = w_n(w_{n-1}(...w_1(w_0))$ where $w_0 = x$. Thus in order to obtain the gradient of $f(x)$, one can either find the gradient of the simpler functions with respect the independent variable or, alternatively, find the gradient of the original function with respect the simpler functions. The first of these different strategies to accumulate terms in the chain rule corresponds to forward AD while the second to backwards AD. Whether to use forwards or backwards AD will greatly depend on the nature of the problem. Generally\footnote{The performance of backwards and forwards AD is also impacted by the size of the operation being differentiated.}, given a function  $f\colon\mathbb{R^N}\to\mathbb{R^M}$, forwards AD will be more efficient at computing $\nabla f$ if $N<M$. If, on the other hand, $M<N$, backwards AD is preferred.   

\LB is currently capable of forwards AD through the \jl library \texttt{ForwardDiff.jl} and \texttt{ReverseDiff.jl} respectively. However, only forwards AD is efficiently implemented. Both \texttt{ForwardDiff.jl} and \texttt{ReverseDiff.jl} perform AD by pushing special types of numbers through the original computer program. On the one hand, \texttt{ForwardDiff.jl} uses dual numbers. Dual numbers are expressions of the form $a + \epsilon b$ such that $\epsilon^2 = 0$ but $\epsilon \neq 0$. It can then be shown that given a generic function $f(a + b \epsilon) = f(a) + b\epsilon f'(a)$ (see Sect. 4 of \citet{GuidetoAD} for proof) such that the gradient of the original function can be easily obtained. This means that \texttt{ForwardDiff.jl} imposes little to no constraints on the program it differentiates through\footnote{Since \jl is a typed language, some considerations have to be made about potential type instabilities introduced by the use of dual numbers.}. On the other hand, \texttt{ReverseDiff.jl} uses taping numbers, a special type of numbers that records all the operations the number undergoes, to generate a trace of the basic operations that compose a generic computer program \citep{ADtape}. Given this record, \texttt{ReverseDiff.jl} can then generate an expression for the gradient of the program. Thus \texttt{ReverseDiff.jl}  requires two passes through the original program. First, a forward pass generates the program's operation trace. Then, a backwards pass computes the partial derivatives and accumulates their values as the input is back-propagated. Because of the greater complexity of the algorithm, \texttt{ReverseDiff.jl} imposes strong demands on the computer programs it acts upon. Commonplace computations such as control flow or variable mutation are examples of operations that must be handled carefully.

\begin{figure*} 
    \includegraphics[width=\linewidth]{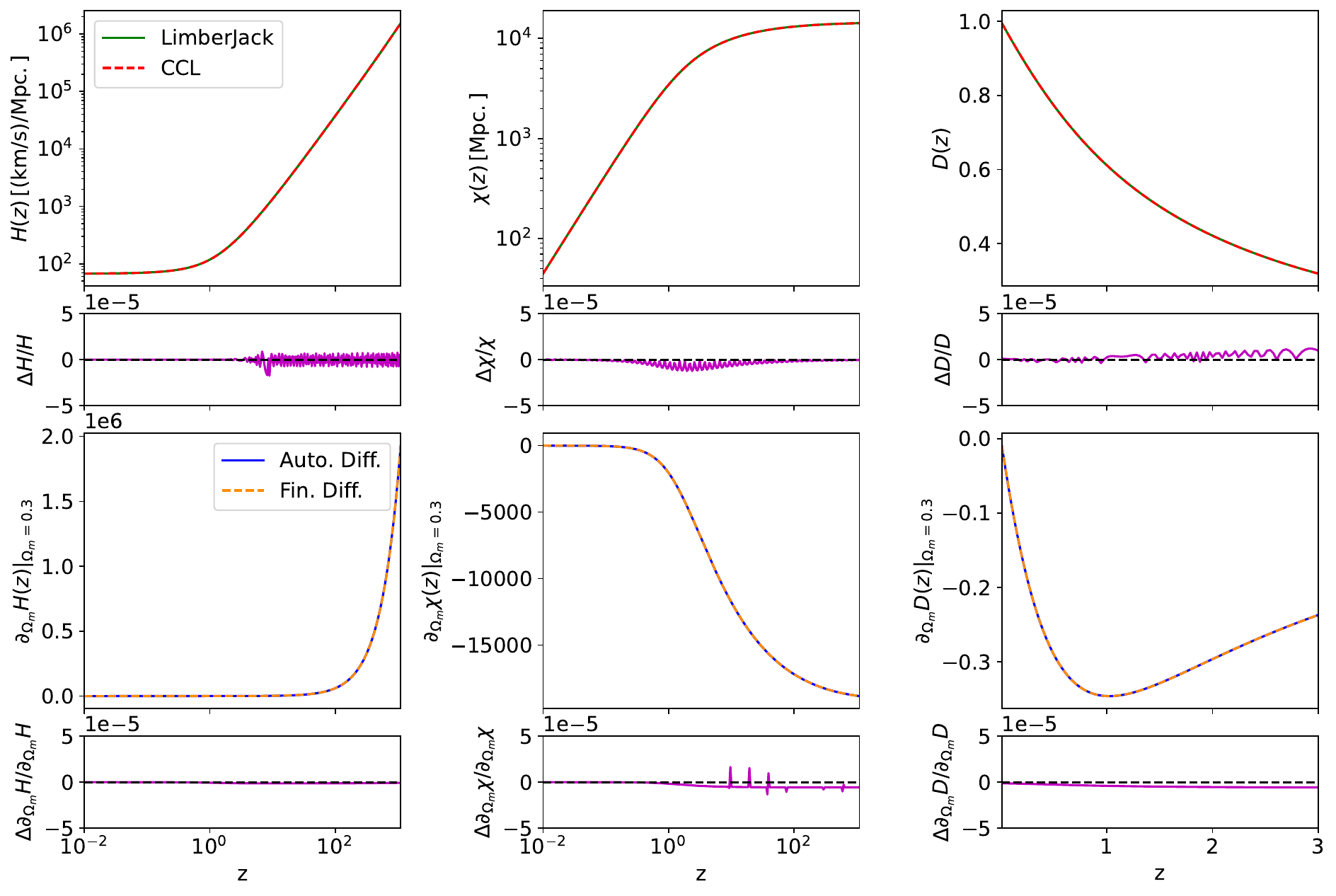} 
    \caption{\textbf{Left column}: Top panel shows a comparison between the \LB (solid green) and the \texttt{CCL} (dashed red) computation of the expansion history between $0<z<1100$. Bottom panel shows a comparison between the derivative of the expansion history with respect to $\Omega_{\rm{m}}$ computed using AD (solid blue) and finite differences (dashed orange) for the same redshift range. \textbf{Middle column}: Top panel shows a comparison between the \LB (solid green) and the \texttt{CCL} (dashed red) computation of the comoving distance between $0<z<1100$. Bottom panel shows a comparison between the derivative of the comoving distance with respect to $\Omega_{\rm{m}}$ computed using AD (solid blue) and finite differences (dashed orange) for the same redshift range. \textbf{Right column}: Top panel shows a comparison between the \LB (solid green) and the \texttt{CCL} (dashed red) computation of the linear growth factor $0<z<3$. Bottom panel shows a comparison between the derivative of the linear growth factor with respect to $\Omega_{\rm{m}}$ computed using AD (solid blue) and finite differences (dashed orange) for the same redshift range.}
   \label{fig:background_acc}
\end{figure*}

The implementation of the expressions discussed in Sect. \ref{Subsect: Angular Power Spectra} within \LB will are thus constrained by the demands of the AD methods used to obtain their gradients. The computation of Eq. \ref{eq:limber} fundamentally involves two different types of quantities, the radial kernels and the matter power spectrum. On the one hand computing the radial kernels boils down to evaluating the expansion history of the Universe, $H(z)$, and its integral over time, i.e. the comoving radial distance, $\chi(z)$. Calculating the expansion history, as given by the $\Lambda$CDM model, amounts to evaluating $H(z) = H_0\sqrt{\Omega_{\rm{m}} (1+z)^3 + \Omega_r (1+z)^4 + (1-\Omega_{\rm{m}}-\Omega_r)}$ on a grid of redshifts where $\Omega_{\rm{m}}$ and $\Omega_r$ are the cosmological matter and radiation density respectively. The radial comoving distance is then obtained integrating over the grid using Simpson numerical integration which is generically compatible with AD methods. In the left and centre columns of Fig. \ref{fig:background_acc} we show a comparison between the \LB and \texttt{CCL} predictions for the expansion history and comoving distance between $0<z<1100$ (top panels). Each panel contains a subpanel where the relative difference between the \texttt{CCL} and \LB predictions is shown. We find that the relative difference between the predictions of \texttt{CCL} and \LB are smaller than $10^{-4}$ at all redshift. Moreover we also show  a comparison for the derivative of said quantities with respect to $\Omega_{\rm{m}}$ when computed by \LB using AD and finite differences for the same redshift range (bottom panels). Similarly, each panel contains a subpanel where the relative difference between the AD and numerical gradient is shown. We find that the relative difference between the two methods is smaller than $10^{-4}$ at all redshift.

On the other hand computing the matter power spectrum in an AD-compatible way is a much more difficult problem. The first and most challenging obstacle is obtaining the linear matter power spectrum. As described in Sect. \ref{Subsect: Angular Power Spectra}, computing the matter power spectrum involves solving a coupled  system of linear differential equations with time- and scale-dependent coefficients. This task has proven a major challenge for most AD environments \citep{JAX-COSMO} as it requires tweaking the lower-level programming languages. This is however not a problem for the \jl AD ecosystem which has a (currently under development) full Boltzmann-Einstein solver, \texttt{Bolt.jl} \citet{Bolt}. While still in its early days, \texttt{Bolt.jl} can provide users with full numerical solutions for the linear power spectrum and its derivatives with respect to the $\Lambda$CDM parameters. \LB can then interface with \texttt{Bolt.jl} to obtain said predictions. 

However, solving the full Boltzmann-Einstein equations is very computationally expensive  (being the bottleneck of most cosmological analyses) even in fast programming languages such as \texttt{C++} (CLASS), \texttt{Fortran} (CAMB) or \jl. Therefore, it is common to look for ways to bypass solving the Boltzmann-Einstein equations in cosmological analyses where speed is important. When the evolution of the matter power spectrum can be assumed to be scale independent, $P(z, k)$ can be computed by constructing a fitting formula that approximates the true value of the matter power spectrum at $z=0$, $P_0(k)$, at an array of scales and then evolving $P_0(k)$ into the past using the linear growth factor. The first commonly applied fitting formula for $P_0(k)$ was the Bardeen-Bond-Kaiser-Szalay \citep[BBKS][]{BBKS} formula which modifies the Harrison-Zeldovich power spectrum \citep{HZ_1, HZ_2, HZ_3} by a transfer function:
\begin{equation}
    P_0(k) = \frac{8 \pi^2}{25} \frac{A_s}{\Omega_{\rm{m}}^2} T^2(k) \frac{k^{n_s}}{H_0^4 k_p^{n_s-1}}\, ,
\end{equation}
where is $A_s$ is the amplitude of the matter power spectrum, $k_p = 0.05 Mpc^{-1}$ and $T(k)$ is the  transfer function for a given set of fixed cosmological parameters. The most popular fitting formula is that of \citep[][E\&H from now on]{EH_1, EH_2}, which follows the same strategy as the BBKS formula of approximating the transfer function but with a more complex expression that includes baryonic effects such as baryonic acoustic oscillations and small-scale power suppression. Thanks to these inclusions the E\&H formula is accurate enough to return unbiased cosmological constraints for the 3x2-pt DES Y1 analysis as shown in \citet{JAX-COSMO}. However, it is important to keep in mind that the E\&H formula will not be accurate enough to analyse future data set as well as some current ones. 

In recent years, a new family of fitting formulae, known as emulators, have grown in popularity. Emulators are computer models (such as neural networks or Gaussian processes) whose weights are optimized to reproduce a target function over a certain domain. The main advantage of emulators over traditional fitting formulae is that they automate the majority of the trial and error process of building an accurate fitting formula. Moreover, they tend to require less knowledge of the physical problem to be constructed. However, emulators require vasts amounts of training data and the resulting models tend to be far larger than traditional fitting formulae. Nonetheless, emulators have found great success in astrophysics in recent years, offering extremely accurate yet affordable approximations to different functions \citep{donald-mccann_textttmatryoshka_2021, Capse}, including the matter power spectrum \citep{CosmoPower, Mootoovaloo}. 

The great advantage of these fitting formulae is that they effectively amount to algebraic expressions through which AD algorithms can easily differentiate as shown in \citet{CosmoPower-JAX} and \citet{Capse}. Combined with their speed, fitting formulae can often be the preferred way to obtain estimates for the primordial power spectrum in gradient-based analyses. For these reasons, \LB is equipped with both a native \jl implementation of the E\&H formula and the recently developed Gaussian process based emulator \texttt{EmuPk} \citep{Mootoovaloo} for the linear matter power spectrum.

The next challenge is to compute the growth factor, $D(z)$, to find the evolution of the linear power spectrum. As discussed in Sect. \ref{Subsect: Angular Power Spectra}, the growth factor is obtained by solving the Jeans equation (see Eq. \ref{eq:jeans_D}) which is an inhomogeneous ordinary differential equation. Differentiating through the solutions of differential equations can be done by writing a recursive numerical scheme to solve the differential equation. These schemes amount to a series of linear operations that update mutating variables. While mutating variables can pose challenges for certain backwards modes of AD, the numerical schemes can otherwise be differentiated through to yield gradients for the solution of a differential equation. Thus, \LB solves the Jeans equation using a second order Runge-Kutta solver. Returning to Fig. \ref{fig:background_acc}, in right column of this figure we show a comparison between the \LB and \texttt{CCL} predictions for the linear growth factor (top panels). We also shown a comparison between the derivatives of \LB's $D(z)$ with respect to $\Omega_{\rm{m}}$ when computed using AD and finite difference (bottom panels) for the redshift range $0<z<3$. Again, each panel contains a subpanel where the relative difference between the two compared quantities is shown. Concerning the growth factor itself, we find that the relative difference between the \texttt{CCL} and \LB predictions are smaller than $10^{-4}$ for all the redshift window. Concerning the derivatives of the growth factor, the relative differences between AD and finite differences are again smaller than $10^{-4}$ for all the redshift window. 

\begin{figure*}
    \centering
    \includegraphics[width=\linewidth]{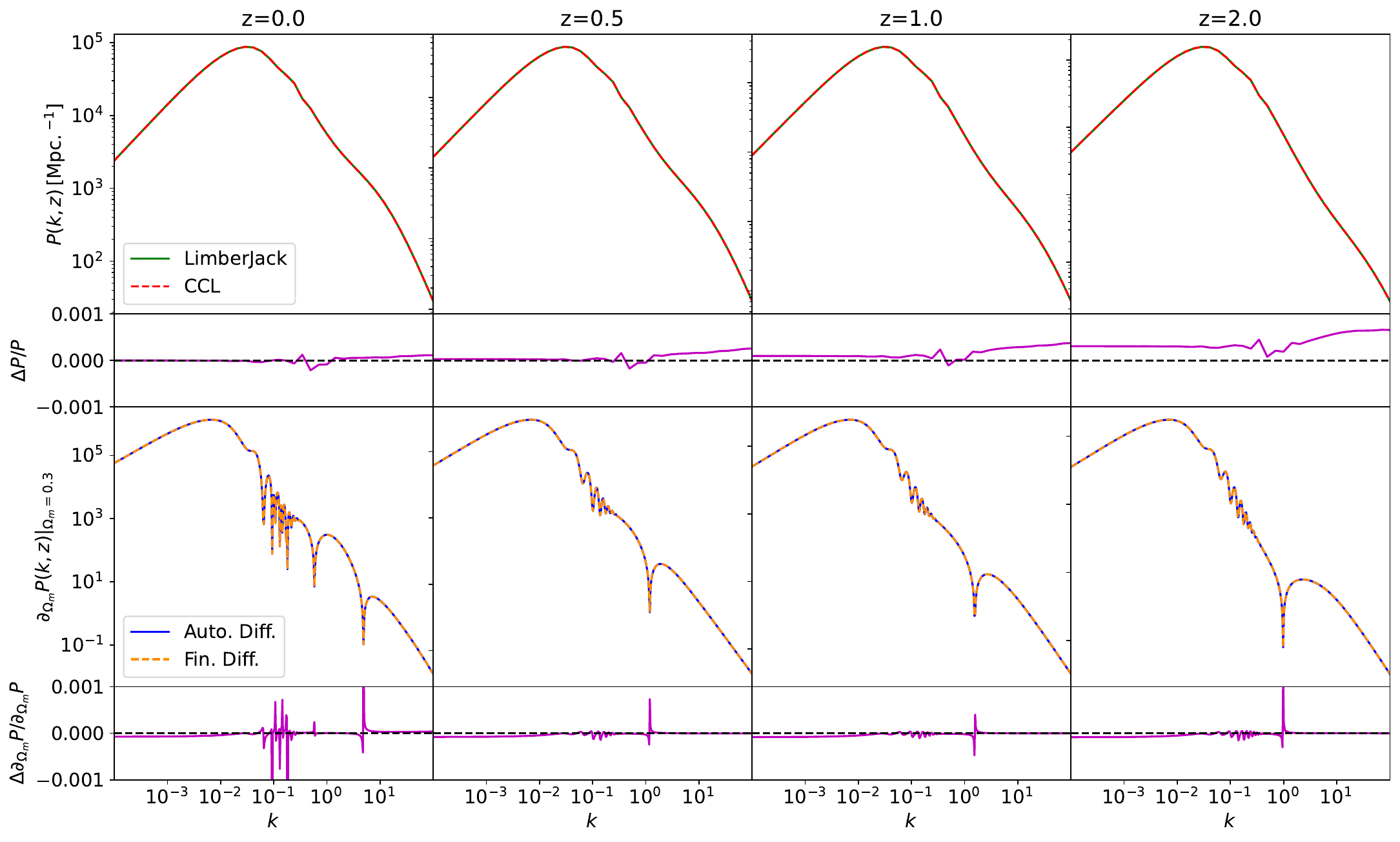}
    \caption{Top panels show a comparison between the \LB (solid green) and the \texttt{CCL} (dashed red) computation of the non-linear matter power spectrum for $z=[0.0, 0.5, 1.0, 2.0]$ and $-4<\log_{10}(k)<2$. Both \LB and \texttt{CCL} used the E\&H formula to compute the linear matter power spectrum. Then, Halofit was used to obtain the non-linear corrections in both cases. Bottom panels shows a comparison between the derivative of the \LB matter power spectra with respect to $\Omega_{\rm{m}}$ computed using AD (solid blue) and finite differences (dashed orange). Each panel contains a subpanel where the relative difference between the two compared quantities is shown.}
    \label{fig:Pk_acc}
\end{figure*}

To compute the necessary non-linear corrections at small scales, \LB is equipped with an auto-differentiable implementation of \texttt{Halofit} as given by \citet{Halofit_1} with revisions of \citet{Halofit_2}. The biggest obstacle to accomplish this is differentiating through the root finding process that occurs within  \texttt{Halofit}. Similarly to solving differential equations, root finding can be differentiated through by writing a recursive numerical scheme of linear operations with mutable variables \citep{Margossian_AD}. The \texttt{Halofit} implementation within \LB uses the secant method for its root finding through which most AD methods can differentiate.  Putting all of these together \LB can perform a fully auto-differentiable computation of the non-linear matter power spectrum at any redshift or scale.  In Fig. \ref{fig:Pk_acc} we show a comparison between the predictions for the non-linear matter power spectrum from \LB and \texttt{CCL} for $z=[0.0, 0.5, 1.0, 2.0] $ and $-2<\log_{10}(k)<4$ (top panels). Both the \texttt{CCL}  and \LB predictions are obtained using the E\&H  linear power spectrum. Then \texttt{Halofit} was used to obtain the non-linear corrections. Moreover we also show a comparison between the derivative of the \LB power spectra with respect to $\Omega_{\rm{m}}$ when computed using AD and finite differences. Finally, each panel contains a subpanel where the relative difference between the two compared quantities is shown. Concerning the power spectra, we find that the relative difference between the \texttt{CCL} and \LB predictions are smaller than $1*10^{-3}$ for all the redshift window. However, we observe that the Halofit implementation of \LB systematically over-predicts the non-linear matter power spectrum at higher redshifts. As we will see this bias will manifest when considering angular power spectra involving the CMB lensing tracer later on but it nonetheless doesn't prevent us from achieving our accuracy goals. In the future, we aim to implement more accurate prescriptions of the non-linear matter power spectrum such as as \texttt{HMCode} \citep{HMCode_og} or \texttt{baccoemu} \citep{BACCO_emulator}. Concerning the derivatives of the same power spectra, the relative differences between AD and finite differences are again smaller than $10^{-3}$ for all the redshift windows. 

The only step left to compute angular power spectra is to bring the radial kernels and the matter power spectrum together and perform the Limber integral (see Eq. \ref{eq:limber}). Similarly to the computation of comoving distances, this is done within \LB by evaluating all the quantities within the integrand at a regular array of logarithmic scales, $log_{10}(k)$, and then performing the numerical integral for the desired multi-poles, $\ell$, using Simpson numerical integration. One small challenge in doing so is finding the corresponding redshift associated with the comoving distance given by the scale and the multipole to evaluate the radial kernels. This inconvenience is a result of \LB defining the radial kernels as functions of redshift instead of comoving distance. Normally, finding the redshift at a given comoving distance would involve a costly root finding process. However, \LB handles this by building an interpolator between redshift and comoving distance and then inverting the interpolator to establish a straight forward comoving distance to redshift map for a given set of cosmological values. 

\begin{figure*}
\centering
\includegraphics[width=\linewidth]{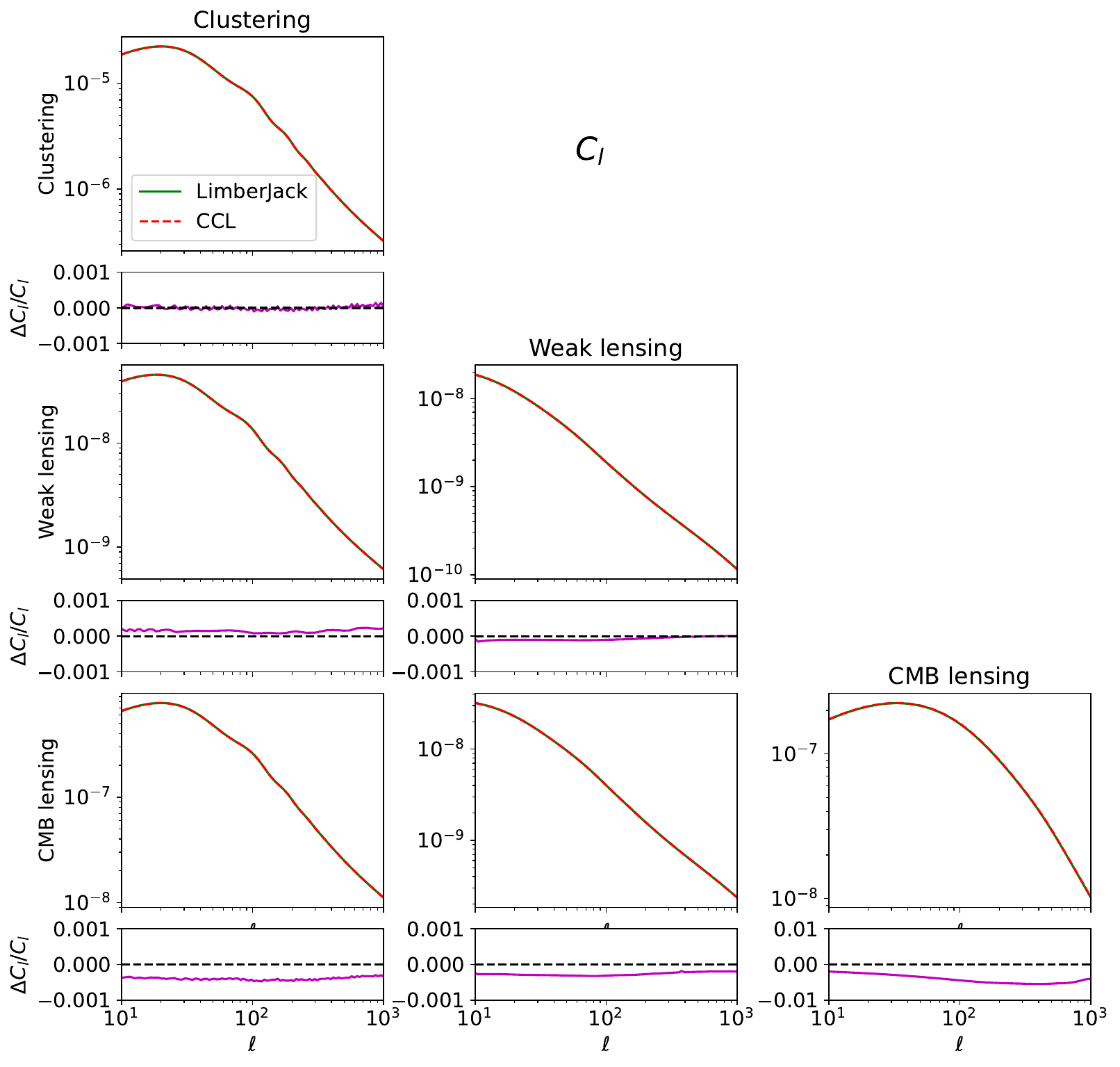} 
\caption{Shows a comparison between the \LB (solid green) and the \texttt{CCL} (dashed red) computation of auto- and cross-correlation angular power spectra of galaxy clustering, weak lensing and CMB lensing tracers for the range of multipoles $1<\log(\ell)<3$. Both \LB and \texttt{CCL} used the E\&H formula to compute the linear matter power spectrum. Then, Halofit was used to obtain the non-linear corrections in both cases.}\label{fig:cls_acc}
\end{figure*}

\begin{figure*}
\includegraphics[width=\linewidth]{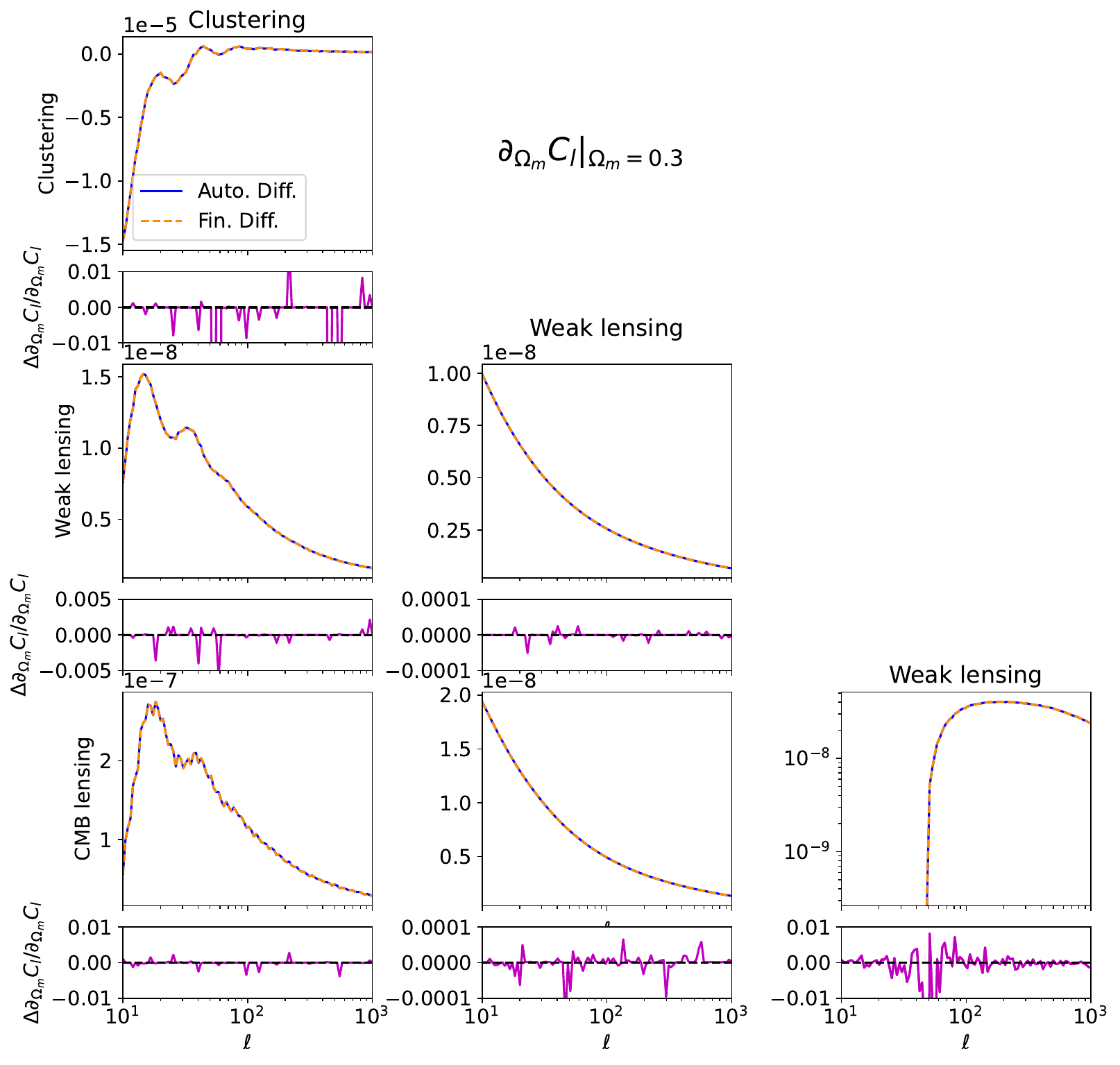} 
\caption{Shows a comparison between the derivatives of the auto- and cross-correlation angular power spectra of galaxy clustering, weak lensing and CMB lensing tracers  with respect to $\Omega_{\rm{m}}$ computed using AD (solid blue) and finite differences (dashed orange)  for the range of multipoles $1<\log(\ell)<3$. In order to compute the gradients of the angula power spectra the E\&H formula was used to compute the linear matter power spectrum. Then, Halofit was used to obtain the non-linear corrections in both cases.}\label{fig:diff_cls_acc}
\end{figure*}

In Fig. \ref{fig:cls_acc} we show a comparison between the predictions of \LB and \texttt{CCL} for the angular power spectra of different types of tracers for $10<\ell<1000$. We consider the auto- and cross-correlations of galaxy clustering, cosmic shear and CMB convergence. In all cases the E\&H formula was paired with \texttt{Halofit} to obtain the matter power spectrum. We also assume a Gaussian redshift distribution centred at $z=0.5$ with a standard deviation $\sigma_z=0.05$, sampled at 1000 evenly-spaced intervals in the range $0<z<2$. We observe that the discrepancy between \LB and \texttt{CCL} is smaller than $10^{-3}$ for all angular power spectra except for the auto-correlation of CMB lensing where the discrpancy is larger but nontheless stays below $10^{-2}$. This is due to the discrepancy in the evolution of the Halofit non-linear matter power spectrum between \LB and \texttt{CCL} shown in Fig. \ref{fig:Pk_acc}. Similarly, in Fig. \ref{fig:diff_cls_acc} we show a comparison for the derivatives of the same quantities with respect to $\Omega_{\rm{m}}$ when computed using AD and finite differences. The derivative of the clustering tracer proved to be extremely sensitive to the resolution of the numerical integration scheme used to normalise the galaxy distributions leading the observed oscillatory behaviour. Despite this, we nonetheless observe a sub-percentage-level agreement between the two approaches in all cases. It also worth remembering that disagreements between AD and finite differences do not necessarily imply a mistake in the AD. Indeed finite differences are more likely to differ from the underlying true derivative given that they are subject to truncation errors.

In this section we have only displayed derivatives of the theoretical predictions of \LB  with respect to $\Omega_{\rm{m}}$. In App. \ref{App: Full Derivatives} we show the derivatives of the non-linear matter power spectrum as well as the auto- and cross-correlation power spectra of galaxy clustering, weak lensing and CMB lensing with respect the five \lcdm parameters for completeness.

\subsection{Gradient-based Samplers} \label{subsect: Gradient-based Samplers}

In the previous sections we described how the computation of angular power spectra (Sect. \ref{Subsect: Angular Power Spectra}) is made compatible with AD methods within \LB (Sect. \ref{Subsect: limeberJack}). In this section we will describe how these newly found gradients can be used to significantly speed up the statistical inference of cosmological parameters, the goal of this paper.

As described in Sect. \ref{Sect: introduction} stochastic inference methods become prohibitively expensive as the number of parameters (dimensions) increases. Gradient-based samplers provide a solution to this problem by using the gradient of the posterior to guide the exploration of the parameter space towards the regions where the probability density of the likelihood is the highest.

Hamiltonian Monte Carlo \citep[HMC, ][]{HMC} is a sampling algorithm that simulates Hamiltonian trajectories on the parameter space to guide the exploration. In order to do so HMC duplicates the parameter space by introducing a set of auxiliary momentum variables ($\boldsymbol{p}$) independent of the likelihood, and sampled from a d-dimensional unit variance Gaussian distribution. The joint distribution of the position (i.e the original parameters of interest $\boldsymbol{\theta}$) and momentum variables is then defined by a Hamiltonian function  which governs the dynamics of the system:

\begin{equation} \label{eq:H}
    H(\boldsymbol{\theta}, \boldsymbol{p}) = T(\boldsymbol{p}) + V(\boldsymbol{\theta}) = \frac{\boldsymbol{p}^2}{2M} - \log(\mathcal{L}(\boldsymbol{\theta})) \, ,
\end{equation}
where $M$ is the mass matrix, a positive-definite matrix that controls the mixing of momenta variables. 

In each iteration of HMC, the Hamiltonian dynamics are simulated using a numerical integration scheme to generate the next sample. This is done using a leapfrog scheme given its symplectic properties that preserve the volume of the phase space along the trajectory. The sample is then accepted or rejected using a Metropolis adjustment based on its Hamiltonian energy. This ensures that the target posterior distribution is obtained upon marginalizing over the momenta variables, i.e: \begin{equation}\label{Eq:canonical_target}
    \mathcal{P}(\boldsymbol{\theta}) \propto \int \exp(-H(\boldsymbol{\theta}, \boldsymbol{p})) \, d\boldsymbol{p} \, . 
\end{equation}

While HMC can dramatically speed up the exploration of large parameter spaces, it also introduces a series of challenges. Namely, the user must find the ideal step size for the numerical solver of the Hamiltonian trajectories, the ideal length of the trajectories and an expression for the mass matrix. The No-U-turns sampler (NUTS) \citep{NUTS} is an HMC algorithm that dynamically tunes these quantities based only on two user provided quantities: the Target Acceptance Probability (TAP) of the Metropolis adjustment and the number of adaptation steps. The fundamental idea behind NUTS is to prevent the sampler from returning to previously explored regions by avoiding U-turns in phase space. This is done by evolving the Hamiltonian trajectories both backwards and forwards in time (known as branching) and stopping when the momenta start to turn on themselves on either branch. This establishes a principled criterion for the length of the Hamiltonian trajectories that maximises the statistical independence of the samples. Moreover, during the adaptation phase the step size of the integrator is tuned such that the TAP is achieved. Finally, NUTS tunes the mass matrix using the variance of the adaptation phase samples such that the covariance matrix of the momenta variables is as close as possible to unit-variance. 

\section{Results} \label{Sect: Results}
\subsection{DES Y1 3x2} \label{Subsect: DESY1}

\begin{figure} 
    \includegraphics[width=\linewidth]{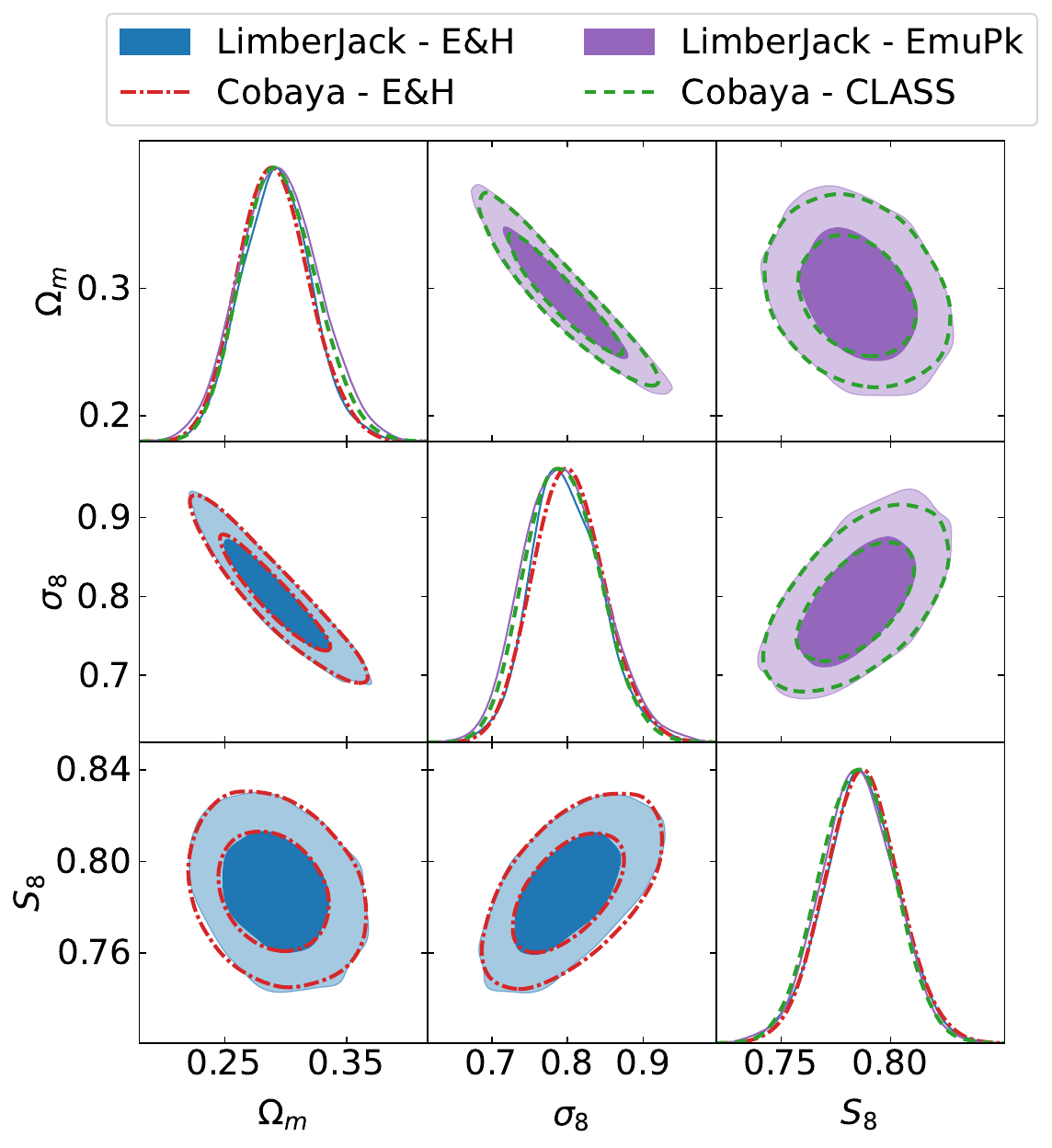} 
    \caption{Shows the marginalised posteriors distributions from the analysis of DES Y1 3x2-pt data for the cosmological parameters $\Omega_{\rm{m}}$, $\sigma_{\rm{8}}$ and $S_{\rm{8}}$. Lower panel compares the \LB (solid blue) and \texttt{CCL} (dashed red) analyses when using the E\&H formula. Upper triangle compares the \LB (solid purple) and \texttt{CCL} (dashed green) analyses when using \texttt{EmuPk} and \texttt{CLASS} respectively. \texttt{Cobaya} contours were obtained using the MH sampler while \LB contours were obtained using the \texttt{NUTS} sampler.}
   \label{fig:DESY1}
\end{figure}

In order to validate the constraints obtained using \LB we first replicate two different analyses based of DES Y1 3x2-pt \citep{DES, DESY1} data done using the well-established Bayesian inference framework $\texttt{Cobaya}$ which employs a Metropolis Hastings (MH) sampler. In the first analysis, both \LB and \texttt{Cobaya} used the E\&H formula to obtain the matter power spectrum. In the second analysis, \texttt{Cobaya} used the $\texttt{CLASS}$ code to obtain the matter power spectrum while \LB used the \texttt{EmuPk} emulator \cite{Mootoovaloo}, which was trained on  $\texttt{CLASS}$. We use the \citet{Garcia-Garcia21} (CGG21 from now on) angular power spectra and covariance matrix of the DES Y1 3x2-pt analysis based of the DES Y1 cosmological catalogue \citet{DESY1} (See Sects. 3.1, 4.1 and 42. of CGG21). The same scale cuts as in CGG21 (see Tab. 4 of CGG21) were applied to the data vector.

With this aim, we make use of the \jl  library for statistical inference \texttt{Turing.jl} \citep{ge2018t}. \texttt{Turing.jl} is a probabilistic programming language (PPL) that allows the user to create statistical models by explicitly writing the relationship between the sampled parameters and the theoretical predictions for the observed data.  \texttt{Turing.jl} uses this information to draw a likelihood density function compatible with the \jl AD infrastructure as long as the internal computations of the model are also compatible with AD. The user can then condition this function on the observations and sample it using a series of sampling back-ends. In this work we make use of the NUTS sampler as implemented in the \jl library \texttt{AbstractHMC.jl} within \texttt{Turing}. Moreover, we choose the AD library \texttt{ForwardDiff.jl} to provide NUTS with the gradient of the likelihood. Note that using forwards AD for statistical inference is sub-optimal since the likelihood is a function that maps a high-dimensional space to a single scalar. In the future, we aim to implement efficient backwards AD in \LB.

\begin{table}
      \caption{Prior distributions for the DES Y1 3x2-pt analysis parameters based on the prior range of \texttt{EmuPk} \citep{Mootoovaloo}}\label{tab:DESY1_priors}
      \centering
      \def\arraystretch{1.2}
      \begin{tabular}{|cc|cc|}
        \hline
        \multicolumn{4}{|c|}{\textbf{DES Y1 3x2-pt Analyses Priors}} \\
        \hline
        Parameter &  Prior & Parameter &  Prior\\  
        \hline 
        \multicolumn{2}{|c|}{\textbf{Cosmology}}               &     \multicolumn{2}{c|}{\textbf{Shear calibration bias }} \\
        $\Omega_{\rm{m}}$  &  $U (0.2, 0.6)$                   &     $m^i $         & $N (0.012, 0.023)$ \\
        $\Omega_{\rm{b}}$  &  $U (0.028, 0.065)$               &     \multicolumn{2}{c|}{\textbf{Galaxy bias }} \\
        $h$                &  $U (N(70, 5), 0.55, 0.91)$       &     $b_g^i $       & $N (0.8, 3.0)$ \\
        $n_{\rm{s}}$       & $U (0.87, 1.07)$                  &     \multicolumn{2}{c|}{\textbf{Intrinsic Alignments}} \\
        $\sigma_{\rm{8}}$  &  $U(0.6, 0.9)$                    &      $A_\mathrm{IA,0} $ & $U(-5.0, 5.0)$ \\
                           &                                   &      $\eta_\mathrm{IA}$  & $U (-5.0, 5.0)$ \\ 
        \multicolumn{2}{|c|}{\textbf{Lens redshift calibration}}& \multicolumn{2}{c|}{\textbf{Sources redshift calibration}} \\
        $\Delta z_1 $  & $N (0.0, 0.007)$                    &     $\Delta z_1 $  & $N (-0.001, 0.016)$ \\ 
        $\Delta z_2 $  & $N (0.0, 0.007)$                    &     $\Delta z_2 $  & $N (-0.019, 0.013)$  \\
        $\Delta z_3 $  & $N (0.0, 0.006)$                   &     $\Delta z_3 $  & $N (0.009, 0.011)$ \\
        $\Delta z_4 $  & $N (0.0, 0.01)$                    &     $\Delta z_4 $  & $N (-0.018, 0.022)$ \\
        $\Delta z_5 $  & $N (0.0, 0.01)$                    &                    &  \\ 
        \hline
      \end{tabular}
\end{table}

The same priors were used in the two cases for both the $\texttt{Cobaya}$ and $\texttt{LimberJack}$ analyses. A summary of the priors can be found in Tab. \ref{tab:DESY1_priors}.  We show the resulting posteriors in Fig. \ref{fig:DESY1}. We observe an excellent agreement between the \LB and \texttt{Cobaya} pipelines regardless of whether the E\&H formula or \texttt{CLASS}/\texttt{EmuPk} is used to obtain the primordial matter power spectrum. 

We compare the performance of the two samplers by looking at the number of effective samples \citep[i.e. number of statistically independent samples in the Markov chain][]{gelman2013bayesian} per number of likelihood calls. This metric is independent of the hardware used to run the analysis and the time taken by \LB and \texttt{Cobaya} to evaluate the likelihood. Thus this metric allows us to look at the improvement purely brought about by using gradient-based inference methods. Analysing the Markov chains of the different samplers we find that \texttt{NUTS} is approximately 1.5 times more efficient than  \texttt{Cobaya} at generating effective samples. However, in order to fairly compare the performance of the two samplers we must take into account that \texttt{NUTS} computes the gradient of the likelihood at every step. Given the 25 parameters of the DES Y1 3x2-pt analysis, evaluating the gradient of the likelihood using \texttt{ForwardDiff.jl} is roughly 5 to 6 times more expensive than evaluating the likelihood itself which is an order of magnitude faster than if we had used finite differences. In order to compensate for this extra cost the efficiency improvement of NUTS over MH should be equal or larger than the relative cost of computing the gradient of the likelihood. Therefore the efficiency improvement of \texttt{NUTS} is not enough to compensate for the cost of computing the gradient in this particular application. 

There exist two avenues to tilt the balance in favour of \texttt{NUTS}. On the one hand, increasing the efficiency of the sampler. On the other hand, reducing the cost of the likelihood gradient. However, the current bottleneck is the efficiency of the sampler as even more specialised AD methods (such as \texttt{Zygote.jl}) would take at least twice the time to compute the likelihood to obtain its gradient. Future works could explore initialising the NUTS mass matrix using the posterior covariance estimate from variational inference algorithms such as \texttt{PathFinder.jl} \citep{PathFinder}. Therefore, in order to showcase the strength of the methods developed on this work we need an application for which the efficiency of gradient-based samplers truly outpaces that of traditional inference methods.

\subsection{Growth Factor Reconstruction} \label{Subsect: GP}

In the previous section we showed the reliability of \LB by reproducing the official DES Y1 3x2-pt analysis. In this section we will showcase how  \LB can be used to perform statistical inference on models outside of the reach of traditional inference methods. In order to do so we perform a model-independent reconstruction of the growth factor using a Gaussian process (GP) with more than a hundred parameters. 

A GP is a collection of random variables (nodes), each of them sampled from a multivariate Gaussian distribution with a non-diagonal covariance \citep{Rasmussen, 1204.2832}. Thus a GP -- $g(\textbf{x})$ where $\textbf{x}$ is a arbitrary vector representing the position of the nodes -- is fully specified by a mean function, $m(\textbf{x}) \equiv \mathbb{E}[g(\textbf{x})]$ -- where $\mathbb{E}[\cdots]$ is the expectation value over the ensemble --  and a covariance function, $k(\textbf{x}, \textbf{x'}) \equiv \mathbb{E}[(g(\textbf{x})-m(\textbf{x}))(g(\textbf{x'})-m(\textbf{x'}))]$. In combination, the mean and covariance functions determine the statistical properties of the random variables that define the family of shapes the GP can take. 

In this work we use a GP composed of 101 nodes equally spaced through the redshift window $0 \le z \le 3$. The mean of the GP is given by the best-fit \lcdm  Planck 2018 \citep[][P18 from now on]{Planck} prediction for $D(z)$. While this choice of mean will bias the reconstruction towards the P18 prediction outside of the data domain, its impact in the data dominated regions is small \citep[see e.g.][for similar examples of this GP behaviour in the literature.]{Ruiz_Zapatero_21, Ruiz_Zapatero_22}. Regarding the covariance matrix, we choose a square exponential covariance function, defined as:
\begin{eqnarray} 
K(\boldsymbol{z},\boldsymbol{z}')=\eta^2\exp\frac{-\mathfrak{D}(\boldsymbol{z},\boldsymbol{z}')}{2l^2} \, ,  \label{eq:QuadExp}
\end{eqnarray}
where $\eta$ is the amplitude of the oscillations around the mean and $l$ is the correlation length between the GP realizations and $\mathfrak{D}(\boldsymbol{z},\boldsymbol{z}')$ is the distance matrix of $\boldsymbol{z}$ and $\boldsymbol{z}'$ given by $\mathfrak{D}_{ij} =  |z_i - z'_j|^2$. This decision was made based on the fact that the square exponential is a computationally inexpensive and infinitely differentiable kernel, appropriate to model smooth fluctuations around the mean of the GP.

Therefore, the growth factor is given by:
\begin{equation} \label{eq:Dz_gp}
    D(\boldsymbol{z}) = D_{\rm{P18}}(\boldsymbol{z}) + L_C(K( \boldsymbol{z}, \boldsymbol{z})) \, \boldsymbol{v} \, ,
\end{equation}
where $\boldsymbol{v}$ is a vector of the GP nodes sampled from a unit variance, diagonal, multivariate normal distribution, $\boldsymbol{z}$ denotes the redshift array at which said nodes sit, $D_{\rm{P18}}(\boldsymbol{z})$  is the P18 growth factor evaluated at $\boldsymbol{z}$ and $L_C(K( \boldsymbol{z}, \boldsymbol{z}))$ is lower-triangular matrix obtained from the Cholesky decomposition of the covariance matrix. A way of interpreting Eq. \eqref{eq:Dz_gp} is as a rotation given by $L_C$ on a vector of white noise $\boldsymbol{v}$ that imposes the correlations of the GP kernel.

This model is similar to the one considered in CGG21, but using a GP to reconstruct the growth factor instead of splines. The reasoning behind the choice of GPs over splines for this work is threefold. First, GPs  offer a well-defined measure of the uncertainty in their predictions which makes assessing the statistical significance of their results straightforward. Second, GPs are not subject to the strict assumptions that  can bias spline reconstructions such as the choice of linear or cubic interpolations. However, it is important to bear in mind that GPs are far from assumption free as their structure is constrained by the properties of the chosen covariance kernel. Nonetheless, as we will show, these assumptions can indeed be neglected when the data is constraining enough. Third, GPs are as differentiable as their covariance matrix kernel \citep{Rasmussen}. The derivative of a GP $g(\boldsymbol{x}) \sim \mathcal{N}(m(\boldsymbol{x}), K(\boldsymbol{x}, \boldsymbol{x}'))$ is another GP given by $\dot{g}(\boldsymbol{x}) \sim \mathcal{N}(\partial_{\boldsymbol{x}}m(\boldsymbol{x}), \partial_{\boldsymbol{x}}\partial_{\boldsymbol{x'}}K(\boldsymbol{x}, \boldsymbol{x}'))$. This is a very desirable feature that will allow us to use growth rate measurements (i.e. measurements of the gradient of the growth factor) to further constrain the reconstructed growth. As it will be shown, the growth rate measurements will highly restrict the evolution of the growth.

\begin{figure} 
    \includegraphics[width=\linewidth]{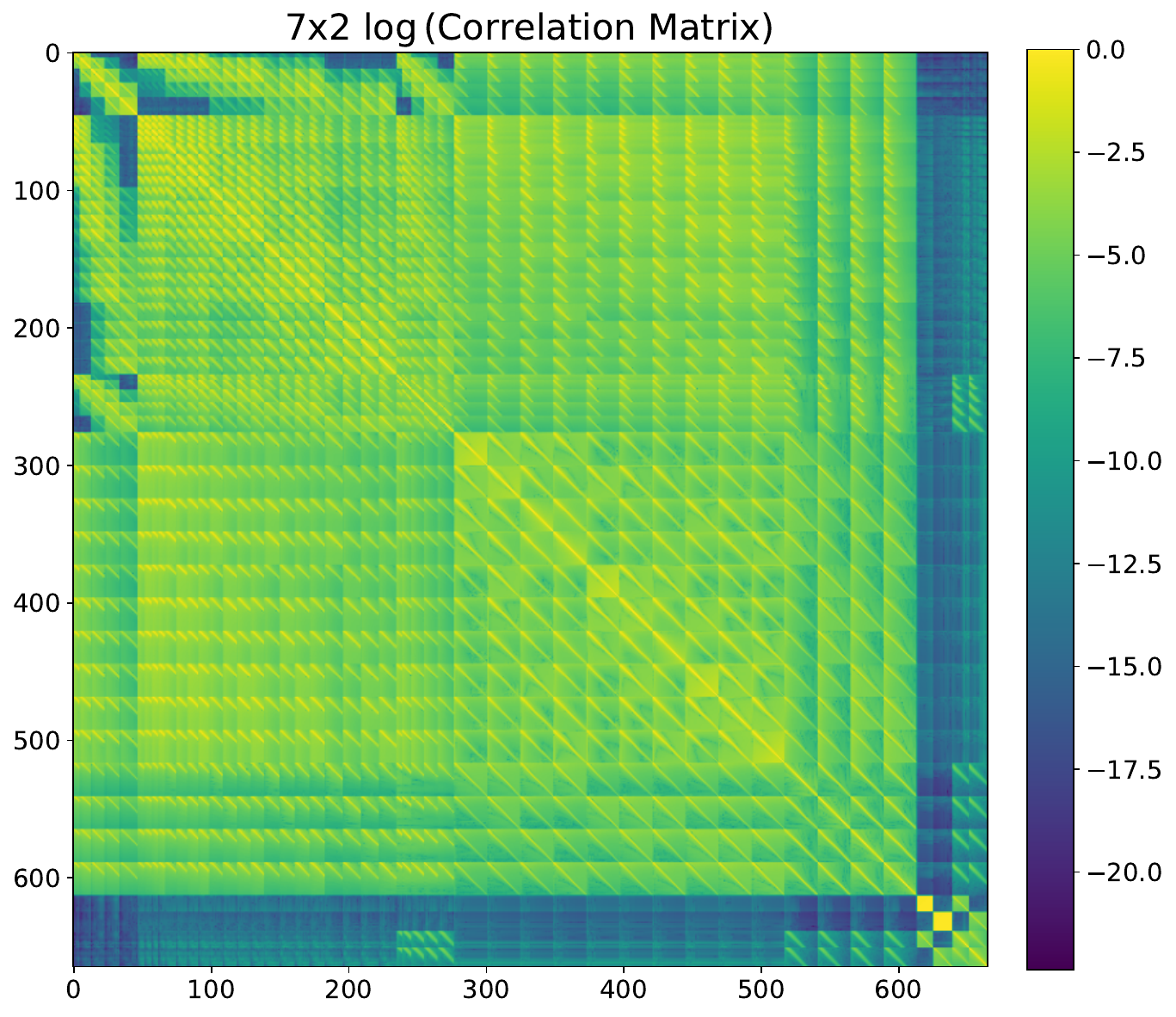} 
    \caption{Shows logarithm of the correlation matrix of the \sxt data set described in Tab. \ref{tab:data_vector}.}
   \label{fig:7x2_cov}
\end{figure}

\begin{figure} 
    \includegraphics[width=\linewidth]{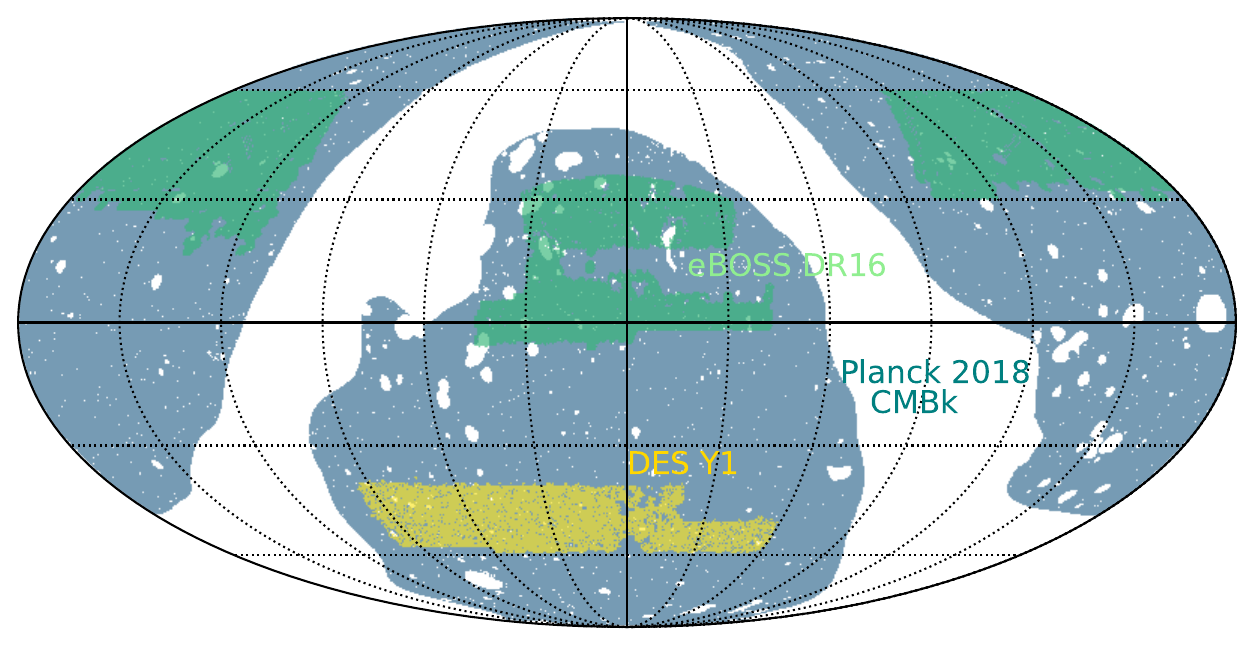} 
    \caption{Shows footprint of the surveys forming the \sxt data vector described in in Tab. \ref{tab:data_vector}.}
   \label{fig:7x2_footprint}
\end{figure}

\begin{figure} 
    \includegraphics[width=\linewidth]{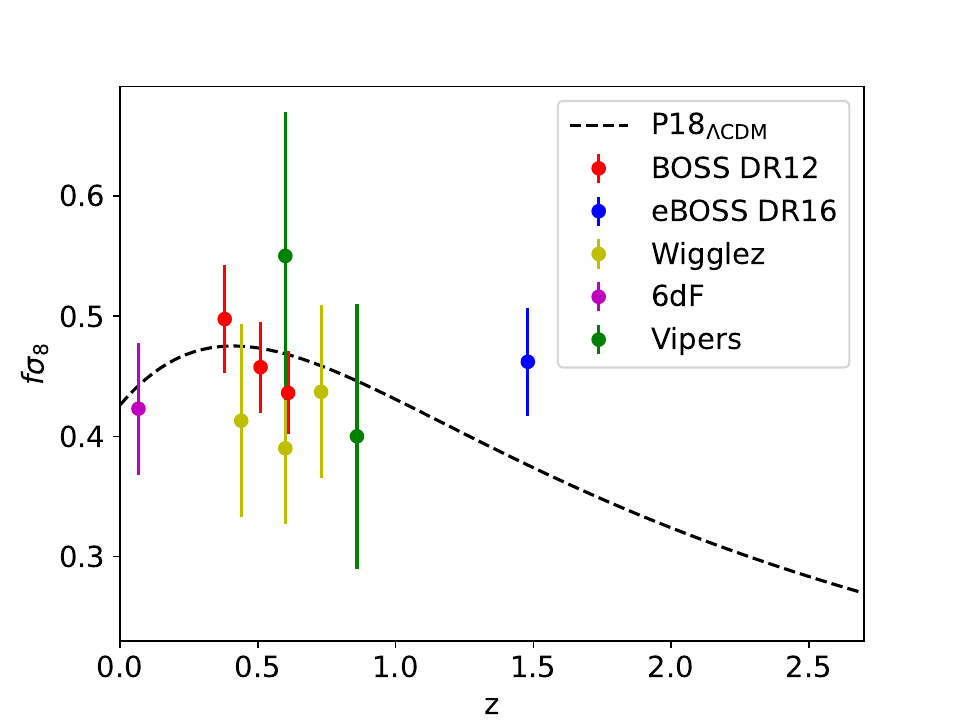} 
    \caption{Shows the $f \! \sigma_{\rm{8}}$ data points from the different surveys used in this work across redshift. Numerical values can be found in Tab. \ref{tab:fs8_data}}
   \label{fig:fs8_data}
\end{figure}

In order to constrain this model we use the south data collection described in CGG21 composed of the 3x2-pt DES Y1 data, the auto-correlation of eBOSS DR16 quasars and the cross-correlation of CMB lensing data with eBOSS DR16 quasars, DES Y1 clustering and DES Y1 weak lensing data. Therefore, our analysis combines a total of 7 two-point statistics which we will refer to as ``\sxt'' hereafter. These particular cross- and auto-correlations correspond to the physical overlap of the different surveys in the sky shown in Fig. \ref{fig:7x2_footprint}. We explicitly list the considered auto- and cross-correlations in Tab. \ref{tab:data_vector}. In this table we can see that the DES Y1 galaxy clustering (GC) data is divided into 5 redshift tomographic bins. The DES Y1 weak lensing (WL) data is divided into 4 redshift tomographic bins. Similarly, the eBOSS DR16 quasar data is divided into 2 redshift tomographic bins. We used the Planck 2018 lensing convergence map. We process all these data following the CGG21 angular power spectra analysis described in Sects. 3.1, 3.2, 3.3, 4.1 and 4.2. Thus we consider a total of 42 different angular power spectra which amount to 665 different data points. The associated covariance matrix of these data was computed using \texttt{Cosmoteka} \citep{Cosmoteka} and it is shown in Fig. \ref{fig:7x2_cov}. The scale cuts considered for each angular power spectrum are listed in the triangle of Tab. \ref{tab:data_vector}.  For a detailed description of these data we refer the reader to Sect. 3 of CGG21. Besides the aforementioned data, we also consider a collection of redshift-space distortion (RSD) measurements by the BOSS DR12 \citep{BOSS12}, eBOSS DR16 \citep{eBOSS16}, Wigglez \citep{Wigglez}, 6dF \citep{6dF} and VIPERS \citep{Vipers} surveys. This analysis constitutes the first combination of all these different data, dramatically improving the constraints by constraining the evolution of the growth factor. We summarise these data in Tab. \ref{tab:fs8_data} and plot it in Fig. \ref{fig:fs8_data}. For a full description of these data we refer the reader to \citet{Ruiz_Zapatero_21}.

\begin{table*}
     \caption{Shows the auto- and cross-correlations that form the \sxt data vector. The first row below the title shows the names of the data sets, namely DES Y1 galaxy clustering (GC) and weak lensing (WL), eBOSS DR16 quasars (eBOSS) and the Planck 2018 CMB lensing convergence map (Planck CMB$\kappa$). The row below shows the tomographic bins of each survey. The triangle table shows the scale cuts in angular scales ($\ell$) applied to each angular power spectrum. In total, the \sxt data vector added to 665 data points.  }\label{tab:data_vector}
      \centering
      \def\arraystretch{1.2}
      \begin{tabular}{|c|c|c|c|c|c|c|c|c|c|c|c|}
        \hline
        \multicolumn{12}{|c|}{\textbf{\textbf{\sxt Data Vector}}} \\
        \hline
        \multicolumn{5}{|c|}{DES Y1 GC} & \multicolumn{4}{c|}{DES Y1 WL} & \multicolumn{2}{c|}{eBOSS} & Planck CMB$\kappa$ \\
        \hline
        0 & 1 & 2 & 3 & 4 & 0 & 1 & 2 & 3 & 0 & 1 & 0 \\ 
        \hline
        \hline
        $0\!-\!145$ &  &  &  &  & $0\!-\!145$ & $0\!-\!145$ & $0\!-\!145$ & $0\!-\!145$ &  &  & $30\!-\!145$ \\
        \cline{1-12}
        \multicolumn{1}{c|}{} & $0\!-\!225$ &  &  &  & $0\!-\!225$ & $0\!-\!225$ & $0\!-\!225$ & $0\!-\!225$&  &  & $30\!-\!225$ \\
        \cline{2-12} 
        \multicolumn{2}{c|}{} & $0\!-\!298$ &  &  & $0\!-\!298$ & $0\!-\!298$ & $0\!-\!298$ & $0\!-\!298$ &  &  & $30\!-\!298$ \\
        \cline{3-12}
        \multicolumn{3}{c|}{} & $0\!-\!371$ &  & $0\!-\!371$ & $0\!-\!371$ & $0\!-\!371$ & $0\!-\!371$ &  &  & $30\!-\!371$ \\
        \cline{4-12}
        \multicolumn{4}{c|}{} & $0\!-\!435$ & $0\!-\!435$ & $0\!-\!435$ & $0\!-\!435$ & $0\!-\!435$ &  &  & $30\!-\!435$ \\
        \cline{5-12}
        \multicolumn{5}{c|}{} & $30\!-\!2000$ & $30\!-\!2000$ & $30\!-\!2000$ & $30\!-\!2000$ &  &  & $30\!-\!2000$ \\
        \cline{6-12}
        \multicolumn{6}{c|}{} & $30\!-\!2000$ & $30\!-\!2000$ & $30\!-\!2000$ &  &  & $30\!-\!2000$ \\
        \cline{7-12}
        \multicolumn{7}{c|}{} & $30\!-\!2000$ & $30\!-\!2000$ &  &  & $30\!-\!2000$ \\
        \cline{8-12}
        \multicolumn{8}{c|}{} & $30\!-\!2000$ &  &  & $30\!-\!2000$ \\
        \cline{9-12}
        \multicolumn{9}{c|}{} & $79\!-\!590$ &  & $79\!-\!590$ \\
        \cline{10-12}
        \multicolumn{10}{c|}{} & $79\!-\!764$ & $79\!-\!764$ \\
        \cline{11-12}
      \end{tabular}
\end{table*}

These data allow us to constraint the growth factor and its derivative across redshift. In order to do so we relate $f\sigma_{\rm{8}}(z)$ with the reconstructed $D(z)$ by
\begin{equation} \label{eq:fs8}
    f\sigma_{\rm{8}}(z) = - \sigma_{8 \, P18} (1+z) \frac{ \partial D(z)}{\partial z} \, , 
\end{equation}
where $\sigma_{8 \, P18}$ is the $\sigma_{8}$ of the \lcdm P18 cosmology and $ D(z)$ is given by Eq. \ref{eq:Dz_gp}.
The main difference in Eq. \ref{eq:fs8} with respect to the \lcdm model is that $\sigma_{8}$ is kept fixed. This is because in our \lcdm analysis we defined $D(z=0)=1.0$ such that the amplitude of the growth factor varies by $\sigma_{\rm{8}}$. Unlike in \lcdm, in our GP model the amplitude of the growth factor is not fixed, but varies with each GP realisation. Therefore, the role of $\sigma_{\rm{8}}$ would be completely degenerate with the amplitude of the $z=0$ GP node and hence is kept fixed.  

\begin{table} 
\centering
\caption{Lists the RSD data used in this analysis.}\label{tab:fs8_data}
\begin{tabular}{|p{4cm}|p{1.5cm}|p{1.5cm}|}
 \hline
Data set &  Redshifts & Data Points\\
 \hline

BOSS DR12 \cite{BOSS12} & 0.38 - 0.61 & 3\\
eBOSS DR16 \cite{eBOSS16} &  1.48  & 1 \\ 
Wigglez \cite{Wigglez} &  0.44 - 0.73  & 3 \\
VIPERS \cite{Vipers} &  0.60 - 0.86   & 2 \\
6dF \cite{6dF} &  0.067  & 1 \\
\hline
\end{tabular}
\end{table}

Once we have drawn an expression for $f\sigma_{\rm{8}}(z)$ in our GP model, the next question is how to evaluate it. While this might seem straight forward, the gradient of the growth factor is extremely sensitive to numerical errors  which can lead to non-physical predictions for $f\sigma_{\rm{8}}(z)$. One solution to this problem would be to compute the gradient analytically using the properties of GPs as discussed above. However, in practice evaluating the derivative of the GP analytically involves treating said derivative as a second GP with a cross-covariance matrix with the original GP. Computing and inverting the cross-correlation matrix between the original GP and its derivative at every step of the Markov chain for the number of nodes used in this analysis proved too computationally expensive.  

The solution found was to use the noise properties of the GP to interpolate the GP to a finer grid such that a finite difference can be taken to obtain the gradient to enough precision. Effectively, this is done by applying a Wiener filter to the original GP where the kernel used for the filtering is the covariance matrix of the GP.  In this scenario the Wiener filter is just given by $W(\boldsymbol{z}^*,\boldsymbol{z}) = K(\boldsymbol{z}^*, \boldsymbol{z}^*) K^{-1}(\boldsymbol{z}, \boldsymbol{z})$ where $ \boldsymbol{z}^*$ is the finer nodes array of redshifts. The resampled growth factor is then given by $D(\boldsymbol{z}^*) = W(\boldsymbol{z}^*, \boldsymbol{z}) D(\boldsymbol{z})$. While this approach also involves inverting a matrix, this is just the covariance matrix of the GP which is a much cheaper operation. 

Having discussed our model it is time to discuss how theory and observations are brought together to constrain the model. We built a Gaussian likelihood assuming the RSD and angular power spectra measurements are completely uncorrelated such that the final likelihood is the product of the likelihoods of the individual data types. This is a fair assumption given that the RSD data is mostly located in the northern hemisphere whereas the angular power power spectra are located further South, resulting in a small overlap between the surveys. We then derived constraints for the GP and cosmological parameters by applying Bayes theorem using the priors displayed in Tab. \ref{tab:priors}.

Note that the GP hyperparameters $l$ and $\eta$ were kept fixed. On the one hand, fixing the hyperparameters avoids the volume effects on the posteriors expected from including these parameters.  On the other hand, fixing the hyperparameters to a particular set of values will introduce certain biases in the final reconstruction of the growth factor specially in the regions where data is sparse. For example, fixing the correlation length is expected to induce spurious oscillations in the reconstructed function outside of the data range. Nonetheless the decision to keep both parameters fixed was made as freeing them introduces severe non-Gaussian degeneracies in the posteriors which pose a challenge even to gradient-based methods such as HMC or NUTS. This is due to their hierarchical relationship with the GP nodes. Future studies that wish to undertake a completely model independent reconstruction will need to explore inference methods specifically tailored to deal with such non-Gaussianity such as Riemannian HMC  \citep{RHMC}.

The number of parameters listed in Tab. \ref{tab:priors} adds up to 128, a dimensionality which requires gradient-based samplers. For this reason, we employed a Gibbs sampling set up where GP parameters were sampled by one NUTS sampler, keeping all other parameters fixed, and then the rest of the parameters (cosmology and nuisances) were sampled by their own NUTS sampler keeping the GP parameters fixed to their last sample. Moreover, the chains were started at the DES Y1 \lcdm best-fit cosmology with the GP nodes starting from zero. The Gibbs scheme combined with the starting point proved to greatly increase the efficiency of the NUTS adapting phase in this high dimensional space. 

In total four analyses were run. First, we ran two \lcdm analyses, one of the \sxt data and another of \sxt + RSD data. In these analyses, we used the \lcdm model to predict the growth factor as opposed to doing a GP reconstruction. Then the same two analyses were ran but performing the GP reconstruction. In addition to these, we also rerun the CGG21 analysis of \sxt data using the priors listed in Tab. \ref{tab:priors} in order to be able to compare our results against the \texttt{Cobaya} pipeline. 

We also study the sampling efficiency in each of these four cases. A summary of our analysis can be found in Tab. \ref{tab:ESS}. Starting with the \lcdm, we find that the effective sample size per calls of the likelihood of the NUTS algorithm with the \sxt data is once again approximately 1.5 times larger than when using the MH algorithm. Remarkably, we found that computing the gradient of the likelihood in this analysis, using AD, was roughly 5 times more expensive than computing the likelihood itself, the same relative cost as the DES Y1 likelihood gradient. Adding RSD data to the \lcdm analysis results in virtually the same effective sample size per calls of the likelihood when using the NUTS sampler. This is to be expected given that fact that the posteriors remain Guassian and the number of dimensions is unchanged.  Looking at the GP analyses, the addition of 101 extra new parameters renders the MH algorithm computationally unfeasible. Thus, we cannot directly compare the effective sample sizes per calls of the likelihood of the two samplers. However, we can still draw comparisons with previous analyses. We find that the efficiency of the NUTS sampler when performing a GP reconstruction based on the \sxt data is around half the efficiency of the \lcdm analysis of same data using the MH algorithm. Moreover, if we add RSD data to the analysis we find that the efficiency of the sampler becomes virtually identical to that of the \lcdm analysis of the \sxt data using the MH algorithm. This is due to the RSD data constraining the derivative of the growth and, therefore, putting stronger constraints on the GP nodes. Taking into account the cost of the likelihood gradient, we find that adding an extra 101 free parameters increases the relative cost of the gradient of the likelihood to a factor of $\sim 30$ when using AD. However, using AD to obtain the gradient of the likelihood remains an order of magnitude cheaper than using finite differences despite the extra 101 free parameters. This showcases the favourable scaling of AD methods with the dimensionality of the problem. Hypothetically, an even more favourable scaling could be achievable by implementing an efficient backwards AD algorithm to obtain the gradient. In combination these two achievements managed to produce converged constraints for our GP analyses in O($10^2$) CPU hours. For reference, our \lcdm analysis of \sxt data using the MH sampler took 24 CPU hours to converge. Thus our methods make analyses with a O(100) parameters computationally feasible while leaving ample room for speed ups.

\begin{table}
\caption{Effective sample size per calls of the likelihood for MH and NUTS samplers when analysing \sxt and \sxt+RSD data using both \lcdm model as well as the GP reconstruction.}\label{tab:ESS}
    \centering
    \def\arraystretch{1.2}
    \begin{tabular}{|p{3.cm}|c|c|c|}
    \hline
    \multirow{2}{*}{Analysis} &  \multicolumn{2}{c|}{ESS/ Lkl. calls} & \multirow{2}{*}{$\dfrac{t[\nabla \mathcal{L}]}{t[\mathcal{L}]}$} \\
    \cline{2-3}
    & NUTS & MH &  \\
    \hline
    \sxt$_{\Lambda \rm{CDM}}$       & 0.0105  & 0.0073 &  4.95 \\
    \sxt+RSD$_{\Lambda \rm{CDM}}$   & 0.0980  & -      &  5.12 \\
    \sxt$_{\rm{GP}}$                & 0.0030  & -      &  28.45\\
    \sxt+RSD$_{\rm{GP}}$            & 0.0071  & -      &  28.60\\
    \hline
    \end{tabular}
    \label{tab:my_label}
\end{table}

\begin{table}
    \caption{Prior distributions for parameters considered for the Growth Factor Reconstruction. We sample the cosmological parameters keeping $\sigma_{\rm{8}}$ fixed to avoid degeneracies with the Gaussian process. For the same reason the Gaussian process hyperparameters are also kept fixed.}\label{tab:priors}
      \centering
      \def\arraystretch{1.2}
      \begin{tabular}{|cc|cc|}
        \hline
        \multicolumn{4}{|c|}{\textbf{Growth Factor Reconstruction Priors}} \\
        \hline
        Parameter &  Prior & Parameter &  Prior\\  
        \hline 
        \multicolumn{2}{|c|}{\textbf{Cosmology}}               &    \multicolumn{2}{c|}{\textbf{DES Y1 - Sources redshift cal.}}  \\
        $\Omega_{\rm{m}}$  &  $U (0.2, 0.6)$                   &    $\Delta z_{\rm{wl}}^1 $  & $N(-0.001, 0.016)$  \\
        $\Omega_{\rm{b}}$  &  $U (0.028, 0.065)$               &    $\Delta z_{\rm{wl}}^2 $  & $N(-0.0019, 0.013)$  \\
        $h$                &  $U (N(70, 5), 0.55, 0.91)$       &    $\Delta z_{\rm{wl}}^3 $  & $N(0.009, 0.011)$    \\
        $n_{\rm{s}}$       & $U (0.87, 1.07)$                  &    $\Delta z_{\rm{wl}}^4 $  & $N(-0.018, 0.022)$    \\
        $\sigma_{\rm{8}}$  &  0.81                             &    &   \\
        \multicolumn{2}{|c|}{\textbf{DES Y1 - Lens redshift cal.}} & \multicolumn{2}{c|}{\textbf{ Intrinsic Alignments}} \\
         $\Delta z_{\rm{gc}}^1 $  & $N(0.0, 0.007)$                   &     $A_\mathrm{IA,0} $ & $U(-5, 5)$ \\
         $\Delta z_{\rm{gc}}^2 $  & $N(0.0, 0.007)$                   &     $\eta_\mathrm{IA}$ & $U(-5, 5)$  \\
        $\Delta z_{\rm{gc}}^3 $  & $N(0.0, 0.006)$                    &     \multicolumn{2}{c|}{\textbf{ eBOSS - Galaxy bias}} \\
        $\Delta z_{\rm{gc}}^4 $  & $N(0.0, 0.01)$                     &      $b_{\rm{QSO}}^i $    & $U(0.8, 5.0)$ \\
        \multicolumn{2}{|c|}{\textbf{DES Y1 - Shear calibration bias }}&    \multicolumn{2}{c|}{\textbf{Gaussian Process}}  \\
        $m^i$          & $N(0.012, 0.023)$                       &      $\eta$       &        0.2 \\
        \multicolumn{2}{|c|}{\textbf{DES Y1 - Galaxy bias }}      &      $l$          &        0.3  \\
        $b_{\rm{gc}}^i$         & $U(0.8, 3.0)$                       &      $v^i$          &    $U(N(0,1), -2, 2)$  \\
        \hline
      \end{tabular}
\end{table}

\subsubsection{\texorpdfstring{\lcdm}{LCDM}  Results} \label{sect: LCDM Results}

We start by analysing our data using a traditional \lcdm model where the growth factor is obtained from the \lcdm parameters by solving the Jeans equation for the matter anisotropies. This allows us to establish a frame of reference against which to compare the results of the GP reconstruction. Moreover, it also allows us to ground our analysis by comparing it with previous analyses of the same combination of data in the literature, namely CGG21.

\begin{figure} 
    \includegraphics[width=\linewidth]{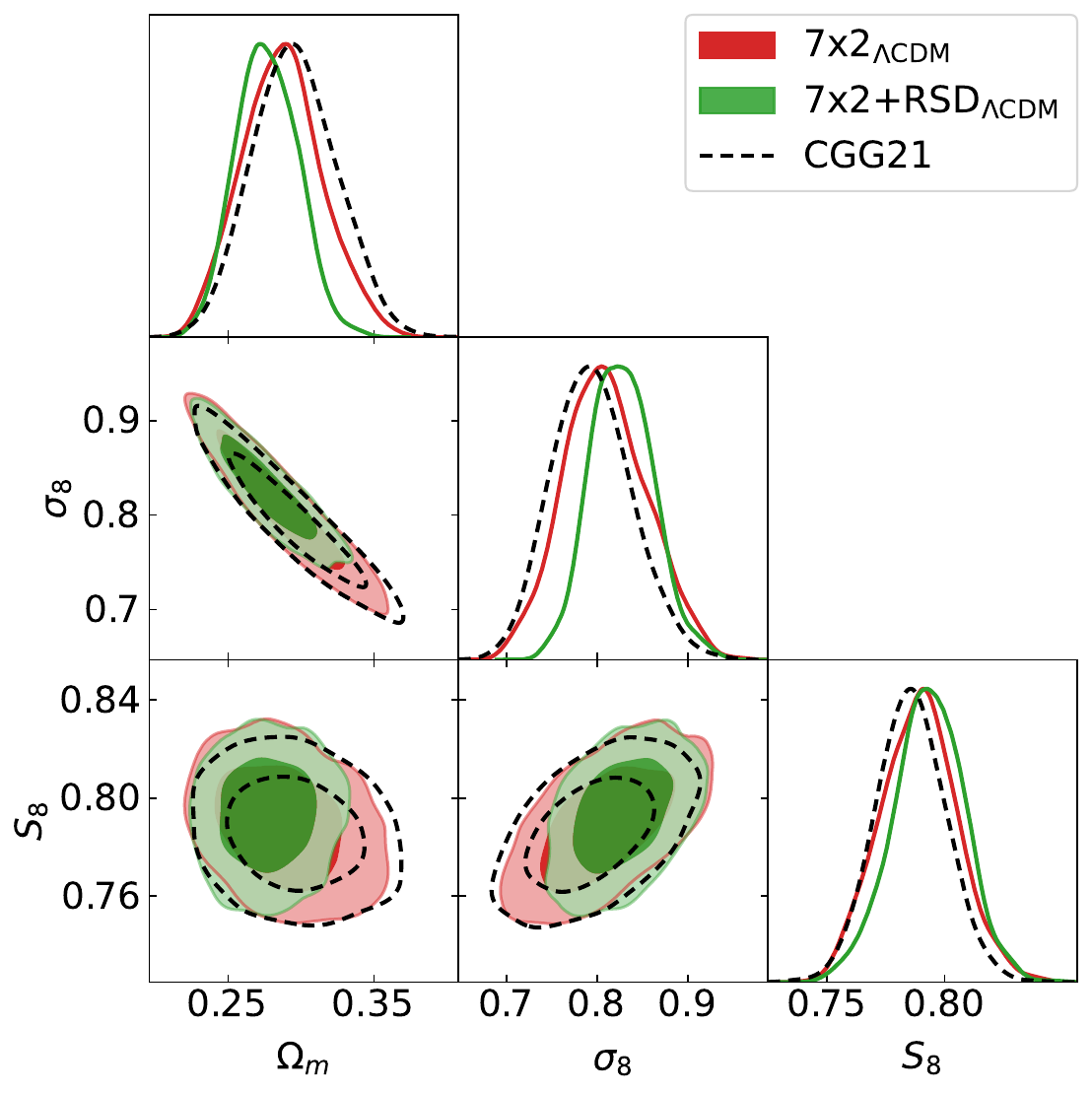} 
    \caption{Shows the 2D and 1D marginal distributions for the parameters $\Omega_{\rm{m}}$, $\sigma_{\rm{8}}$ and $S_{\rm{8}}$ of the \lcdm analysis of \sxt data (red) and \sxt plus RSD data (green). Black dashed contours show the reanalysis of the CGG21 using Tab. \ref{tab:priors}'s priors.}
   \label{fig:7x2_comp}
\end{figure}

In Fig \ref{fig:7x2_comp}, we show the obtained $\Omega_{\rm{m}}$, $\sigma_{\rm{8}}$ and $S_{\rm{8}}$ posteriors when performing a \lcdm analysis of the \sxt data set with and without including additional RSD data (green and red contours respectively). We also show the contours obtained by the CGG21 pipeline when analysing the \sxt data using black dashed lines. Our \lcdm analysis of the \sxt data found $\Omega_{\rm{m}} = 0.287 \pm 0.027$, $\sigma_{\rm{8}} = 0.809 \pm 0.045$ and $S_{\rm{8}} = 0.789 \pm 0.016$. The reanalysis of CGG21 using the priors shown in Tab. \ref{tab:priors} found  $\Omega_{\rm{m}} = 0.296 \pm 0.028$, $\sigma_{\rm{8}} = 0.794 \ 0.043$ and $S_{\rm{8}} = 0.786 \pm 0.015$. Thus we can observe that the constraints produced by the \LB pipeline presented in this work are completely consistent with the results of the \texttt{Cobaya} pipeline of CGG21. Combining the \sxt with the RSD data we found $\Omega_{\rm{m}} = 0.277 \pm 0.021$, $\sigma_{\rm{8}} = 0.827 \pm 0.034$ and $S_{\rm{8}} = 0.793 \pm 0.015$. The addition of RSD data improved the constraints on $\Omega_{\rm{m}}$ and $\sigma_{\rm{8}}$ by 20\%. However, the constraints on $S_{\rm{8}}$ were largely unaffected. Regardless of whether or not RSD data are included, our results are in 2 $\sigma$ disagreement with the P18 results which found $S_{\rm{8}} = 0.832 \pm 0.013$ \citep[See Tab. 1 of][]{Planck}. The full numerical posteriors of the \LB \lcdm analyses of \sxt  and \sxt plus RSD data can be found in the first two columns of Tab. \ref{tab:posteriors}.

\subsubsection{GP Results} \label{sect: GP Results}

We are now in a position to start discussing the GP reconstruction of the growth factor. We start by considering how introducing the GP affects the parameter constraints previously discussed in Sect. \ref{sect: LCDM Results}. In order to establish a comparison we derive constraints for $\sigma_{\rm{8}} \equiv \sigma_{\rm{8}}(z=0)$ and $S_{\rm{8}} \equiv S_{\rm{8}}(z=0)$ from the GP reconstruction of the growth factor. From Eq. \ref{eq:fs8}, we can see that within the GP model:
\begin{eqnarray}
    \sigma_{\rm{8}} = \sigma_{\rm{8}}^{\rm{P18}} D(z=0) \,, \\ 
    S_{\rm{8}} = \sigma_{\rm{8}}^{\rm{P18}} D(z=0) \sqrt{\Omega_{\rm{m}}/0.3} \, ,
\end{eqnarray}
where $D(z)$ is given by \ref{eq:Dz_gp}. 

\begin{figure} 
    \includegraphics[width=\linewidth]{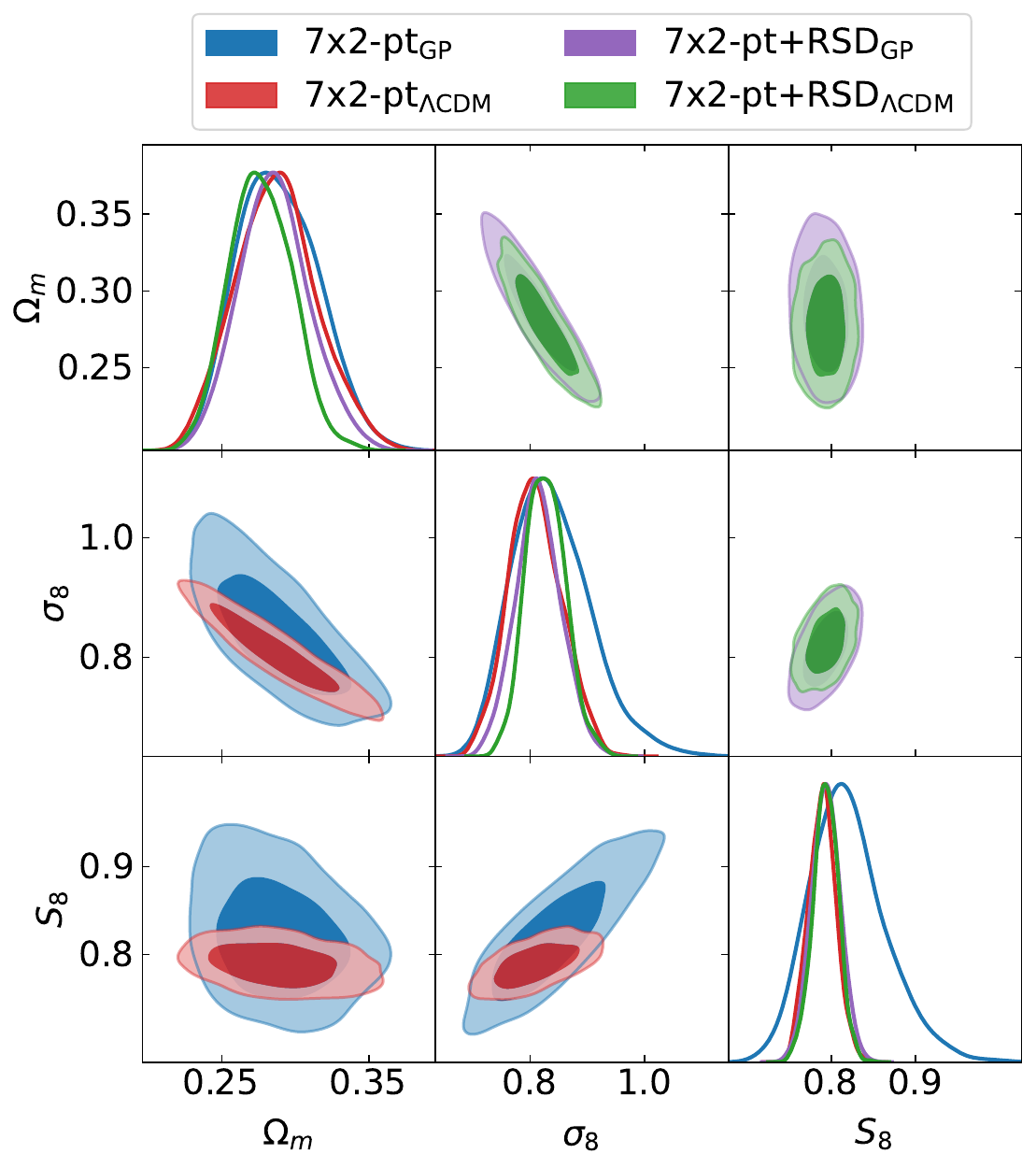} 
    \caption{Lower triangle: shows the 2D marginal distributions for the parameters $\Omega_{\rm{m}}$, $\sigma_{\rm{8}}$ and $S_{\rm{8}}$  of the \lcdm (red) and GP reconstruction (blue) analyses of \sxt data. Upper triangle: shows the 2D distributions for the parameters $\Omega_{\rm{m}}$, $\sigma_{\rm{8}}$ and $S_{\rm{8}}$  of the \lcdm (green) and GP reconstruction (purple) analyses of \sxt data combined with RSD data. 1D Marginals for all the previous analyses are shown along the diagonal.}
   \label{fig:7x2_gp_comp}
\end{figure}

In Fig. \ref{fig:7x2_gp_comp} we show four different set of posteriors for the parameters $\Omega_{\rm{m}}$, $\sigma_{\rm{8}}$ and $S_{\rm{8}}$. In the lower triangle we show the contours obtained when using the \lcdm model (red) and the GP model (blue) to analyse the \sxt data. Similarly, in the upper triangle we show the contours obtained when using the \lcdm model (red) and the GP model (blue) to analyse the \sxt data in addition to RSD data. The associated numerical constraints for the GP reconstructions are $\Omega_{\rm{m}} = 0.289 \pm 0.026$, $\sigma_{\rm{8}} = 0.839 \pm 0.067$ and $S_{\rm{8}} = 0.839 \pm 0.067$ when \sxt alone is consider and  $\Omega_{\rm{m}} = 0.286 \pm 0.023$, $\sigma_{\rm{8}} = 0.815 \pm 0.039$ and $S_{\rm{8}} = 0.793 \pm 0.017$ when RSD data is included. The full posteriors can be found in the last two columns of  Tab. \ref{tab:posteriors}. Performing the GP reconstruction of the growth factor greatly reduces the constraining power of the \sxt data. We observe $\sim$30\% wider constraints when using the GP model to analyse the data on average across the three parameters. It is important to note that the overwhelming majority of the impact occurs in the parameters explicitly related to the growth factor such as $\sigma_{\rm{8}}$ and $S_{\rm{8}}$ and the linear bias parameters of the different probes. Other parameters such as $h$ or $n_s$ remain virtually unchanged. The $S_{\rm{8}}$ constraint is also centred at a significantly higher value than when performing a \lcdm analysis. This due to the lack of data at $z=0$ to constrain the GP, which is reflected in the large error bar in the constraint. When RSD's are included the degradation in constraining power when performing the GP reconstruction is much smaller, with constraints only being $\sim$10\% wider. Moreover, the $S_{\rm{8}}$ constraint becomes centred at exactly the same value as when performing the \lcdm analysis of the same data. Due the larger error bar, we observe that nonetheless the tension with the P18 \lcdm value drops to 1.7 sigma. 

\begin{figure*} 
    \includegraphics[width=\linewidth]{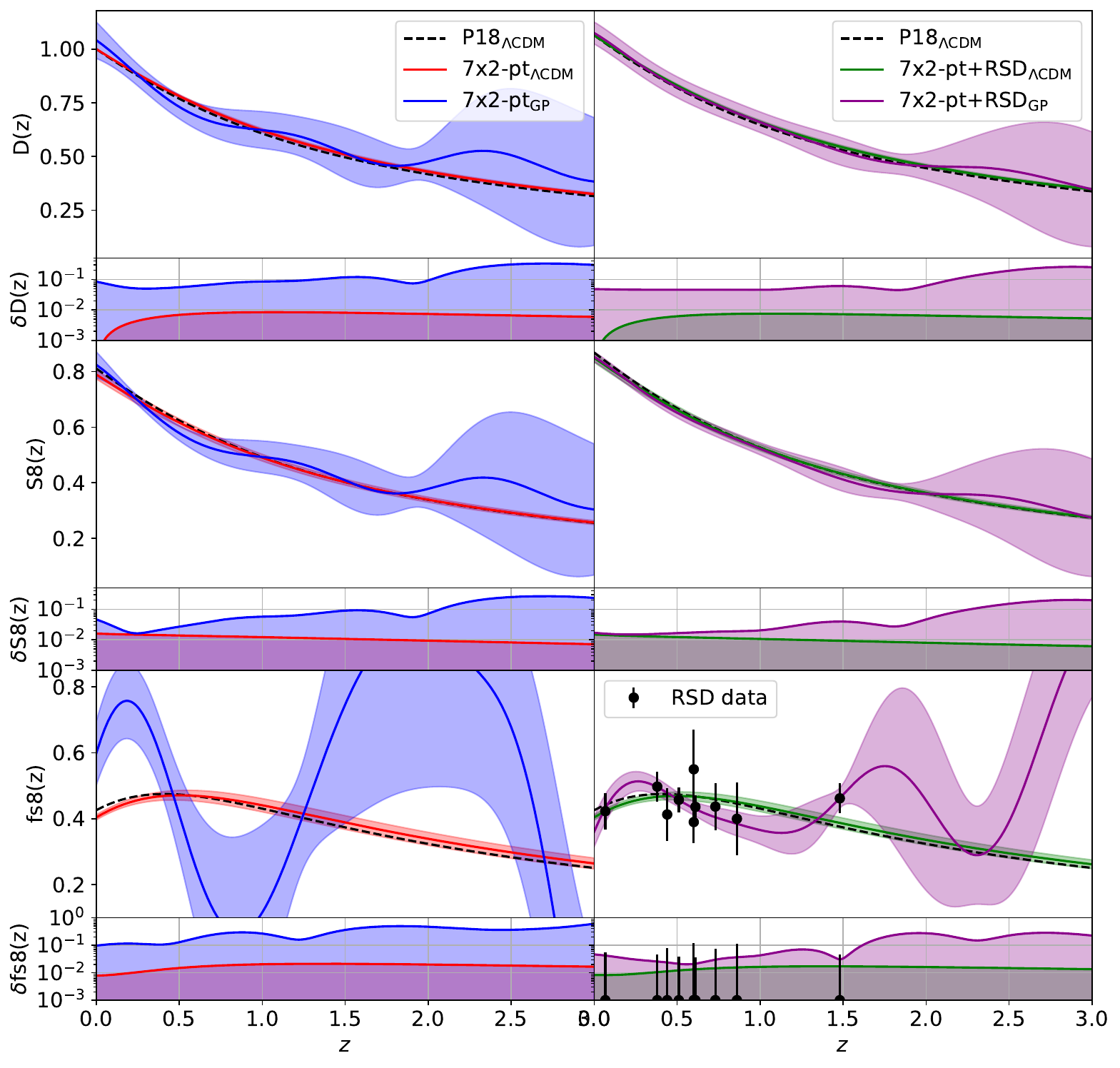} 
    \caption{Shows the evolution across redshift of the reconstructed $D(z)$ (first row), $S_{\rm{8}}(z)$ (second row) and $f\sigma_{\rm{8}}(z)$ (third row). Results based on \sxt data are shown on the left column while results combining \sxt plus RSD data are shown in the right column. In each panel we over plot the prediction of the \lcdm model for the given data as well as the result of the GP reconstruction. Moreover, we also overplot the P18 \lcdm prediction using a black dashed line for reference. Finally, each panel counts with a subpanel where the we show the evolution of the $1\sigma$ confidence intervals of the plotted functions across redshift.}
   \label{fig:gp_comp}
\end{figure*}

In order to understand how the inclusion of RSD data leads to these changes, we have to look at the reconstructed growth factor. In each of the rows of Fig. \ref{fig:gp_comp} we show the constraints obtained for $D(z)$, $S_{\rm{8}}(z)$ and $f\sigma_{\rm{8}}(z)$ (in this order) as functions of redshift. In each panel we compare the constraints obtained when using the \lcdm model against the GP reconstruction. The left-column panels show analyses where only the \sxt data was used while the right column shows the respective analyses when RSD data were included. Moreover, each panel has a subpanel showing the evolution of the $1\sigma$ confidence interval of each function over redshift. 

Let us begin the discussion by focusing on the top panels of Fig. \ref{fig:gp_comp} showing the evolution of the growth factor. In these panels we can see how introducing a GP to reconstruct the growth factor from the data increases the error bars in the predictions of $D(z)$ by one to two orders of magnitude when compared to the \lcdm prediction based on the same data. Including RSD data significantly contributes to constraining $D(z)$ regardless of the model considered. In the case of \lcdm analyses, we observe a 10\% reduction in the standard deviation of $D(z)$ across redshift. The impact of RSD data is even bigger when we instead use a GP to reconstruct $D(z)$. In this case we observe a 20\% to 30\% percent improvement in the constraints depending on the redshift window. Nonetheless, we observe that in all four cases the obtained growth factor is compatible with the P18 prediction at all redshifts. 

It is important to note that the uncertainty of the $D(z)$ \lcdm prediction actually falls to zero at $z=0$ since in this model the growth factor is re-scaled to be precisely one at this redshift. Therefore, by looking at the \lcdm prediction of $D(z)$ we are omitting a large contribution to the uncertainty of the growth factor in this model, its amplitude. Therefore, it is useful to consider quantities such as $S_{\rm{8}}(z)$ that do incorporate the uncertainty in the amplitude of the growth factor in the \lcdm model, encapsulated in the parameter $\sigma_{\rm{8}}$. In the second row of panels of \ref{fig:gp_comp} we show the associated  $S_{\rm{8}}(z)$ constraints for the four scenarios considered. In these panels we can see that the uncertainty in $S_{\rm{8}}(z)$ of the GP reconstruction at low redshift is actually comparable to that of the \lcdm model the uncertainty in the amplitude is taken into account. However, the uncertainty of the GP reconstruction quickly increases once the data becomes sparse as seen in the $D(z)$ panels. 

It is also interesting to note that, as expected, fixing the correlation length of the GP induces oscillations in the reconstructed growth factor and $S_{\rm{8}}(z)$. This is visible in the GP reconstruction of $D(z)$ based on \sxt data. However, including RSD data nullifies these oscillations. In order to understand these behaviours we need to look at the third rows of panels of Fig. \ref{fig:gp_comp}. In these panels we show the associated $f\sigma_{\rm{8}}(z)$ predictions as a function of redshift. In the left panel we can see how the oscillations on the growth factor induced by our assumptions on the GP hyperparameters result in non-physical predictions for $f\sigma_{\rm{8}}(z)$ despite the growth factor itself being perfectly compatible with the P18 \lcdm prediction. Including RSD data solves this problem by directly constraining the possible values of $f\sigma_{\rm{8}}(z)$ given the current data. These constraints on $f\sigma_{\rm{8}}(z)$ translate into strict demands for the evolution of the gradient of the growth factor. Therefore, when RSD data are included, we can see how the oscillatory behaviour present in the reconstruction from only \sxt data disappears. This means that when the data is constraining enough the GP reconstruction remains unbiased by the assumptions introduced by fixing the hyperparameters. 

\begin{figure} 
    \includegraphics[width=\linewidth]{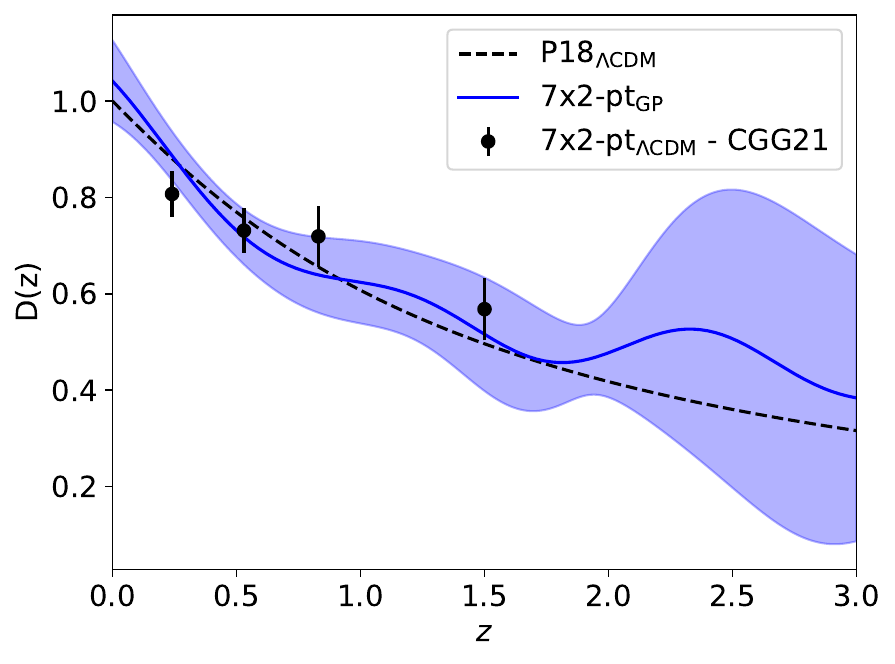} 
    \caption{Shows a comparison between the reconstructed growth factor in this work using a GP (blue) and the CGG21 reconstruction based on splines (black dots) of the same \sxt data. We also overplot the P18 \lcdm prediction for the growth factor (black dashed line) for reference.}
   \label{fig:carlos_comp}
\end{figure}

Finally it is relevant to compare the growth factor reconstruction performed in this work using GPs against the reconstruction performed by CGG21 from the same data using splines. In Fig. \ref{fig:carlos_comp} we can observe that our GP reconstruction stands in good statistical agreement with the four constraints on  $S_{\rm{8}}(z)$ at $z=[0.24, 0.53, 0.83, 1.5]$ obtained by CGG21. This reinforces the notion that these different model-agnostic approaches are data-driven, and not biased by their own individual assumptions.

\begin{table*}
     \caption{Posterior distributions of the Growth Factor Reconstruction analyses. Error bars denote the $1\sigma$ confidence interval.}\label{tab:posteriors}
      \centering
      \def\arraystretch{1.4}
      \begin{tabular}{|c|c|c|c|c|}
        \hline
        \multicolumn{5}{|c|}{\textbf{Growth Factor Reconstruction Posteriors}} \\
        \hline
        Parameter &  \sxt$_{\Lambda \rm{CDM}}$ & \sxt+RSD$_{\Lambda \rm{CDM}}$ &  \sxt$_{\rm{GP}}$ & \sxt+RSD$_{\rm{GP}}$ \\ 
        \hline
        $ \chi^2 $ & 658 & 645 & 652 & 635 \\
        $ \chi^2/(d-p) $ & 1.03 & 0.99 & 1.21 & 1.16 \\
        \hline
    $ \Omega_m $ & $ 0.287^{+0.026}_{-0.030} $ & $ 0.277\pm 0.021 $ & $ 0.289^{+0.025}_{-0.031} $ & $ 0.286^{+0.022}_{-0.026} $ \\
    $ \Omega_b $ & $ 0.0450^{+0.0086}_{-0.012} $ & $ 0.0433^{+0.0084}_{-0.012} $ & $ 0.0448^{+0.0082}_{-0.012} $ & $ 0.0444^{+0.0086}_{-0.012} $ \\
    $ h $ & $ 0.725^{+0.042}_{-0.050} $ & $ 0.733\pm 0.039 $ & $ 0.726\pm 0.040 $ & $ 0.731\pm 0.040 $ \\
    $ n_s $ & $ 0.960\pm 0.056 $ & $ 0.965\pm 0.059 $ & $ 0.961^{+0.058}_{-0.069} $ & $ 0.948^{+0.054}_{-0.078} $ \\
    $ \sigma_{\rm{8}} $ & $0.809^{+0.043}_{-0.051}$ & $0.827^{+0.033}_{-0.037}$ & $0.839^{+0.056}_{-0.079}$ & $0.815\pm 0.039$ \\
    $ S_8 $ & $ 0.789\pm 0.016 $ & $ 0.793^{+0.016}_{-0.014} $ & $ 0.820^{+0.036}_{-0.052} $ & $ 0.793\pm 0.017 $ \\
    $ b_{gc}^1 $ & $ 1.46\pm 0.10 $ & $ 1.418\pm 0.082 $ & $ 1.46\pm 0.10 $ & $ 1.444\pm 0.090 $ \\
    $ b_{gc}^2 $ & $ 1.77\pm 0.11 $ & $ 1.729\pm 0.084 $ & $ 1.84^{+0.12}_{-0.13} $ & $ 1.771^{+0.094}_{-0.11} $ \\
    $ b_{gc}^3 $ & $ 1.75\pm 0.10 $ & $ 1.705\pm 0.078 $ & $ 1.87\pm 0.14 $ & $ 1.752^{+0.092}_{-0.11} $ \\
    $ b_{gc}^4 $ & $ 2.13\pm 0.12 $ & $ 2.081\pm 0.095 $ & $ 2.29^{+0.18}_{-0.26} $ & $ 2.13^{+0.11}_{-0.14} $ \\
    $ b_{gc}^5 $ & $ 2.19\pm 0.13 $ & $ 2.14\pm 0.11 $ & $ 2.31^{+0.22}_{-0.34} $ & $ 2.18^{+0.13}_{-0.15} $ \\
    $ \Delta z_{gc}^1 $ & $ 0.0031\pm 0.0067 $ & $ 0.0040\pm 0.0066 $ & $ 0.0027\pm 0.0067 $ & $ 0.0039\pm 0.0068 $ \\
    $ \Delta z_{gc}^2 $ & $ 0.0029^{+0.0076}_{-0.0066} $ & $ 0.0030\pm 0.0067 $ & $ 0.0016\pm 0.0067 $ & $ 0.0025\pm 0.0066 $ \\
    $ \Delta z_{gc}^3 $ & $ 0.0013\pm 0.0056 $ & $ 0.0006\pm 0.0060 $ & $ 0.0004\pm 0.0057 $ & $ 0.0007\pm 0.0056 $ \\
    $ \Delta z_{gc}^4 $ & $ 0.0025^{+0.0085}_{-0.0097} $ & $ 0.0028\pm 0.0096 $ & $ 0.0019\pm 0.0095 $ & $ 0.0024\pm 0.0090 $ \\
    $ \Delta z_{gc}^5 $ & $ -0.0013\pm 0.0095 $ & $ -0.001\pm 0.010 $ & $ -0.0011\pm 0.0099 $ & $ -0.0010\pm 0.0099 $ \\
    $ m_{wl}^1 $ & $ 0.020\pm 0.023 $ & $ 0.022\pm 0.023 $ & $ 0.016\pm 0.022 $ & $ 0.022\pm 0.022 $ \\
    $ m_{wl}^2 $ & $ 0.014\pm 0.021 $ & $ 0.009\pm 0.022 $ & $ 0.012\pm 0.022 $ & $ 0.009\pm 0.022 $ \\
    $ m_{wl}^3 $ & $ 0.011^{+0.022}_{-0.020} $ & $ 0.009\pm 0.021 $ & $ 0.012\pm 0.020 $ & $ 0.009\pm 0.020 $ \\
    $ m_{wl}^4 $ & $ 0.003\pm 0.022 $ & $ 0.006\pm 0.020 $ & $ 0.007\pm 0.021 $ & $ 0.006\pm 0.021 $ \\
    $ \Delta z_{wl}^1 $ & $0.006\pm 0.012$ & $0.009\pm 0.013$ & $0.003\pm 0.013$ & $0.009\pm 0.013$ \\
    $ \Delta z_{wl}^2 $ & $-0.030\pm 0.010$ & $-0.035\pm 0.011$ & $-0.031\pm 0.011$ & $-0.035\pm 0.011$ \\
    $ \Delta z_{wl}^3 $ & $0.0064\pm 0.0099$ & $0.007\pm 0.010$ & $0.0093\pm 0.0099$ & $0.0070\pm 0.0096$ \\
    $ \Delta z_{wl}^4 $ & $-0.025^{+0.021}_{-0.018}$ & $-0.025\pm 0.019$ & $-0.020\pm 0.020$ & $-0.024\pm 0.018$ \\
    $ b_{QSO}^1 $ & $ 2.24^{+0.19}_{-0.22} $ & $ 2.18\pm 0.17 $ & $ 2.18^{+0.29}_{-0.44} $ & $ 2.26^{+0.23}_{-0.26} $ \\
    $ b_{QSO}^2 $ & $ 2.42\pm 0.22 $ & $ 2.36\pm 0.19 $ & $ 2.28^{+0.32}_{-0.42} $ & $ 2.48^{+0.28}_{-0.33} $ \\
    $ A_{IA} $ & $ 0.28^{+0.16}_{-0.19} $ & $ 0.31\pm 0.19 $ & $ 0.28^{+0.17}_{-0.19} $ & $ 0.31^{+0.17}_{-0.20} $ \\
    $ \alpha_{IA} $ & $ 0.2\pm 2.3 $ & $ 0.1\pm 2.5 $ & $ -0.2^{+2.5}_{-3.2} $ & $ 0.0\pm 2.5 $ \\
    \hline
      \end{tabular}
\end{table*}

\section{Conclusions} \label{Sect: Conclusions}

In this work we have presented \LB, an auto-differentiable cosmological code to compute angular power spectra fully written in \jl. The goal of \LB is to enable the use of gradient-based inference methods in cosmological analyses. \LB core strength's are:
\begin{itemize}
    \item \textbf{Auto-Differentiablility}: Every step in between the input cosmological parameters and the output theoretical prediction in \LB is compatible with \jl's auto-differentiation (AD) libraries \texttt{ForwardDiff.jl} and \texttt{ReverseDiff.jl}. These methods result in gradients up to an order of magnitude faster than when using finite differences. This is the key feature that makes the use of gradient-based inference methods possible with \LB 
    \item \textbf{Speed}: \LB is equipped with a native implementation of the matter power spectrum emulator \texttt{EmuPk} \citep{Mootoovaloo} which makes it orders of magnitude faster than \texttt{CLASS}.
    \item \textbf{Accuracy}: The \LB theoretical predictions have been thoroughly tested against those of the well-established cosmological code $\texttt{CCL}$, as well as the quality of the AD gradients against the more costly finite difference results.  
    \item \textbf{Interoperability}: Thanks to its modular structure, \LB can be easily interfaced with other \jl libraries to increase its capabilities. For example \LB can be interfaced with the \texttt{Bolt.jl} library \citep{Bolt} to gain access to the first auto-differentiable Boltzmann code.
\end{itemize}

Furthermore, we presented two examples of how \LB can be employed to perform present and future cosmological analyses using the gradient-based inference algorithm  NUTS \citep{NUTS}, a self-tuning formulation of the Hamiltonian Monte Carlo algorithm. In the first example we reproduced two analyses of DES Y1 3x2-point data performed using the  well established $\texttt{Cobaya}$ library to ensure the reliability of \LB. \LB's constraints were found virtually identical to those obtained with $\texttt{Cobaya}$ regardless of whether the matter power spectrum was computed using the Eisenstein and Hu \citep{EH_1, EH_2} formula or \texttt{EmuPk} \citep{Mootoovaloo} to emulate \texttt{CLASS} \citep{Class}. Moreover, the NUTS sampler proved to be 1.5 times more efficient (measured as effective samples per likelihood call) than the MH sampler used by Cobaya. However, this improvement in efficiency proved to be not enough to compensate for the cost of computing the gradient of the likelihood despite using AD methods. Further work is necessary to determine the point at which gradient-based inference methods out-weight the cost of computing the likelihood gradient in angular power spectra analyses. 

In the second example we showcased the unique capabilities of \LB by performing a Gaussian process (GP) reconstruction of the growth factor across redshift adding to a total of 128 parameters. In order to constrain this model we employed a combination of DES Y1 galaxy clustering and weak lensing data, eBOSS QSO’s and CMB lensing (referred to as \sxt in the text) as well as a collection of the latest RSD measurements of $f\sigma_{\rm{8}}$. We started by considering a \lcdm analysis of the aforementioned data to establish a baseline for the GP reconstruction. Our \lcdm analysis of the \sxt data found $S_{\rm{8}} = 0.789 \pm 0.016$. Adding the RSD data yielded $S_{\rm{8}} = 0.793 \pm 0.015$. Regardless of whether or not RSD data are included, our results are in $2\sigma$ disagreement with the Planck 2018 results which found $S_{\rm{8}} = 0.832 \pm 0.013$ \citep[See Tab. 1 of][]{Planck}. Performing the GP reconstruction instead yielded $S_{\rm{8}} = 0.839 \pm 0.067$ when \sxt alone is considered and  $S_{\rm{8}} = 0.793 \pm 0.017$ when RSD data is included. Looking at the reconstructed growth factor we observed a reasonable agreement between the GP and the Planck 2018 \lcdm prediction regardless of the data combination used. However, including RSD data significantly smoothed the reconstruction of the growth factor, disfavouring large oscillations. Moreover, it improved the constraints on the reconstructed growth factor by 20\% on average across redshift. This stresses the importance of including RSD data in future cosmological analyses, specially given the up coming DESI survey \citep{DESI}. In terms of sampling efficiency, our GP analysis of \sxt+RSD data using the gradient-based NUTS sampler managed to achieve the same sampling efficiency as our reference \lcdm analysis of \sxt data. Moreover, \LB's AD methods reduced the cost of the likelihood's gradient by an order magnitude with respect finite differences. In combination these two achievements made a previously unfeasible analysis computationally possible, taking O($10^2$) CPU hours to reach convergence.

Auto-differentiable and gradient-based inference methods will play a crucial role speeding up future cosmological analyses as well as enabling entirely new science. For instance, analyses of multiple auto- and cross-correlations between stage-IV surveys may contain up to a hundred free parameters. This will be the case even when performing traditional \lcdm analyses with minimalistic modelling of systematics simply due to the large number of tomographic bins that will be involved. Future surveys will, however, provide unprecedented measurements of small scales. In order to fit these scales and further our understanding of non-linear cosmology, more complex modelling of baryonic effects will have to be included. Similarly, the constraining power provided by the new data will enable analyses with beyond \lcdm physics as well as model-agnostic reconstructions such as the one presented in CGG21 and this work which will have a similar effect in the number of free parameters. In addition to this, gradient-based inference methods are already indispensable to undertake field level inference cosmology \citep{Lavaux_and_leclercq, Bayer_vs, Biwei_seljak} and they will become more so in the future. 

Finally, while \LB is already a fully functional tool, there are several avenues for future improvement:
\begin{enumerate}
    \item \textbf{Improved predictions}: the methods currently implemented in \LB provide enough accuracy to analyse DES Y1 data. However, these methods will need to be improved in order to analyse DES Y3 data or future data sets such as LSST. Here we present a non-exhaustive list of possible extensions:
    \begin{itemize}
        \item Non-linear corrections to the matter power spectrum beyond the Halofit formula. 
        \item Small scale baryonic effects on the galaxy and matter power spectra.
        \item Angular power spectra beyond the Limber approximation \citep{NK5}. 
        \item Scale-dependent growth of structure. 
        \item Perturbatory expansion models for the matter-galaxy bias \citep{Lepori}.
    \end{itemize}
    \item \textbf{Parallelization}: currently the threading parallelisation of \LB is suboptimal. This is because the default \jl threading parallelisation scheme does not handle  the shared memory between the threads efficiently enough. At the moment computing resources are best spent running different instances of \LB in parallel as opposed to parallelising one instance. Future works could study alternative parallelisation schemes or manually managing the memory between the threads to improve the multi-core performance of \LB. 
    \item \textbf{GPU's}: \LB currently cannot run on GPU's which are known to significantly speed-up cosmological inference. This is due to \LB performing scalar indexing operations at several points of the computation of angular power spectra.  Future work could study how to bypass these operations and to make \LB comptible with \jl's GPU libraries such as $\texttt{CUDA.jl}$. 
    \item \textbf{Backwards-AD}: currently \LB's preferred AD mode is forward-AD. However, statistical inference preferred AD mode is backwards-AD, specially as the number of parameters increases. Future works could look into making \LB compatible with the latest \jl AD libraries such as $\texttt{Zygote.jl}$ or $\texttt{Enzyme.jl}$ to implement efficient backwards-AD.
\end{enumerate}

\section*{Author contributions}
\begin{itemize}
    \item \textbf{Jaime Ruiz-Zapatero}: Science lead, lead designer and programmer.
    \item \textbf{David Alonso}: Tracers and Limber integral design. Main advisor.
    \item \textbf{Carlos Garc\'ia-Garc\'ia}:  Science advisor and Cobaya analyses.
    \item \textbf{Andrina Nicola}: \texttt{Halofit} code.
    \item \textbf{Arrykrishna Mootoovaloo}: \texttt{EmuPk} integration.
    \item \textbf{James M. Sullivan}:  \texttt{Bolt.jl} integration.
    \item \textbf{Marco Bonici}: Design advisor and benchmarks. 
    \item \textbf{Pedro G. Ferreira}:  Science advisor.
\end{itemize}

\section*{Software Availability}
\LB is a fully open-source project available on \jl's general repository of packages and \href{https://github.com/jaimerzp/LimberJack.jl}{GitHub} \textcolor{blue}{\faGithub}.

\section*{Acknowledgements}
We would like to thank Francois Lanusse and Marius Millea for their helpful comments.
JRZ is supported by an STFC doctoral studentship. 
DA is supported by the Beecroft Trust.
CGG is supported by the Beecroft Trust.
AN is supported through NSF grants AST-1814971 and AST-2108126
AM is funded through Grant 62192 from the John Templeton Foundation to LSST Corporation. 
PGF is supported by STFC and the Beecroft Trust.
JMS is partially supported by a US Department of Energy Office of Science Graduate Student Researcher award.

The opinions expressed in this publication are those of the authors and do not necessarily reflect the views of LSSTC or the John Templeton Foundation. 

For the purpose of Open Access, the authors have applied a CC BY public copyright licence to any Author Accepted Manuscript version arising from this submission.

We made extensive use of the {\tt numpy} \citep{van2011numpy}, {\tt scipy} \citep{scipy_virtnamen}, {\tt astropy} \citep{astropy_2013, astropy_2018}, {\tt healpy} \citep{Zonca2019}, {\tt GetDist} \citep{getdist}, and {\tt matplotlib} \citep{Hunter_2007} python packages. We also make use of the \texttt{Julia} packages {\tt ForwardDiff.jl} \citep{ForwardDiff},  {\tt AdvancedHMC.jl}  \citep{AHMC} and {\tt Turing.jl} \citep{ge2018t}.

\appendix
\section{Full Derivatives} \label{App: Full Derivatives}
In this section we show the derivatives of the most relevant predictions of LimberJack with respect to the five \lcdm parameters $\Omega_{\rm{m}}$, $\Omega_{\rm{b}}$, $h$, $\sigma_{\rm{8}}$ and $n_{\rm{s}}$. 

\begin{figure*} 
    \includegraphics[width=\linewidth]{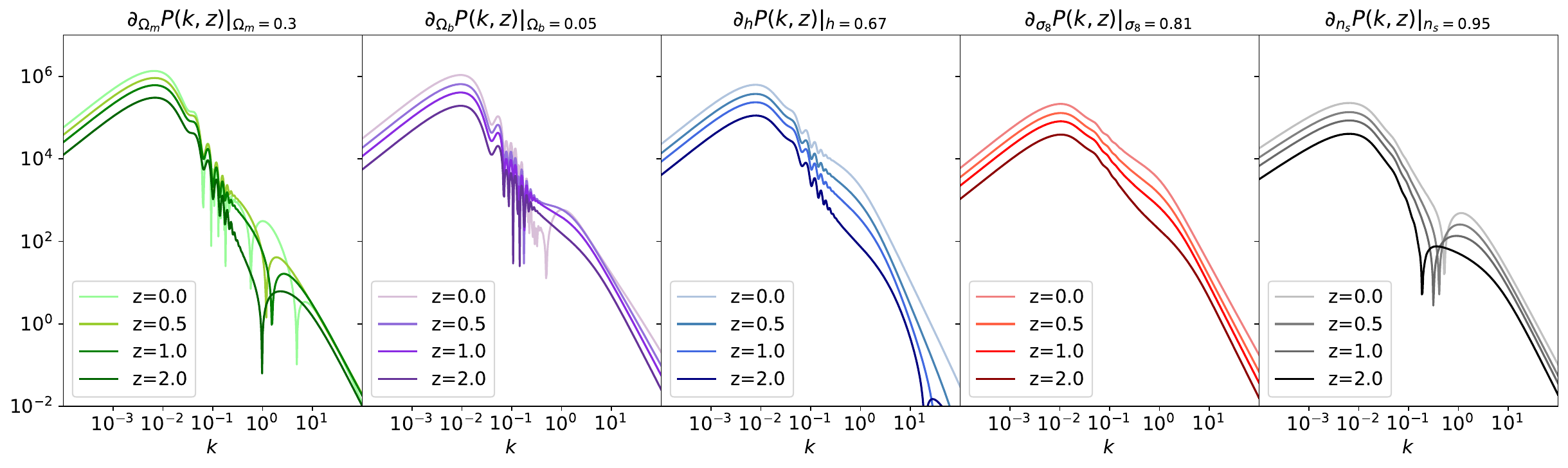} 
    \caption{Shows the derivatives of the non-linear matter power spectrum $P(k,z)$ with respect to the five \lcdm parameters $\Omega_{\rm{m}}$, $\Omega_{\rm{b}}$, $h$, $\sigma_{\rm{8}}$ and $n_{\rm{s}}$. The linear matter power spectrum was computed using the E\&H formula. The non-linear corrections were computed using the Halofit formula. In each panel the we overplot the value of the corresponding derivative at $z=[0.0, 0.5, 1.0, 2.0]$ where light colours correspond to lower redshifts. All the derivatives where evaluated at $\Omega_{\rm{m}}=0.30$, $\Omega_{\rm{b}}=0.05$, $h=0.67$, $\sigma_{\rm{8}}=0.81$ and $n_{\rm{s}}=0.95$}
   \label{fig:Pk_diffs}
\end{figure*}

In  FIg. \ref{fig:Pk_diffs} we show the derivatives of the non-linear matter power spectrum where the linear matter power spectrum was computed using the E\&H formula. In each panel the we overplot the value of the corresponding derivative at $z=[0.0, 0.5, 1.0, 2.0]$ where light colours correspond to lower redshifts. As we can see, the derivatives with respect $\Omega_{\rm{m}}$ and $\Omega_{\rm{b}}$ of $P(k,z)$ are highly oscillatory. On the other hand, the rest of the derivatives are smoother. Moreover, we can also observe that the value of the derivatives decreases for all five parameters as the redshift at which they are computed increases.  

\begin{figure*} 
    \includegraphics[width=\linewidth]{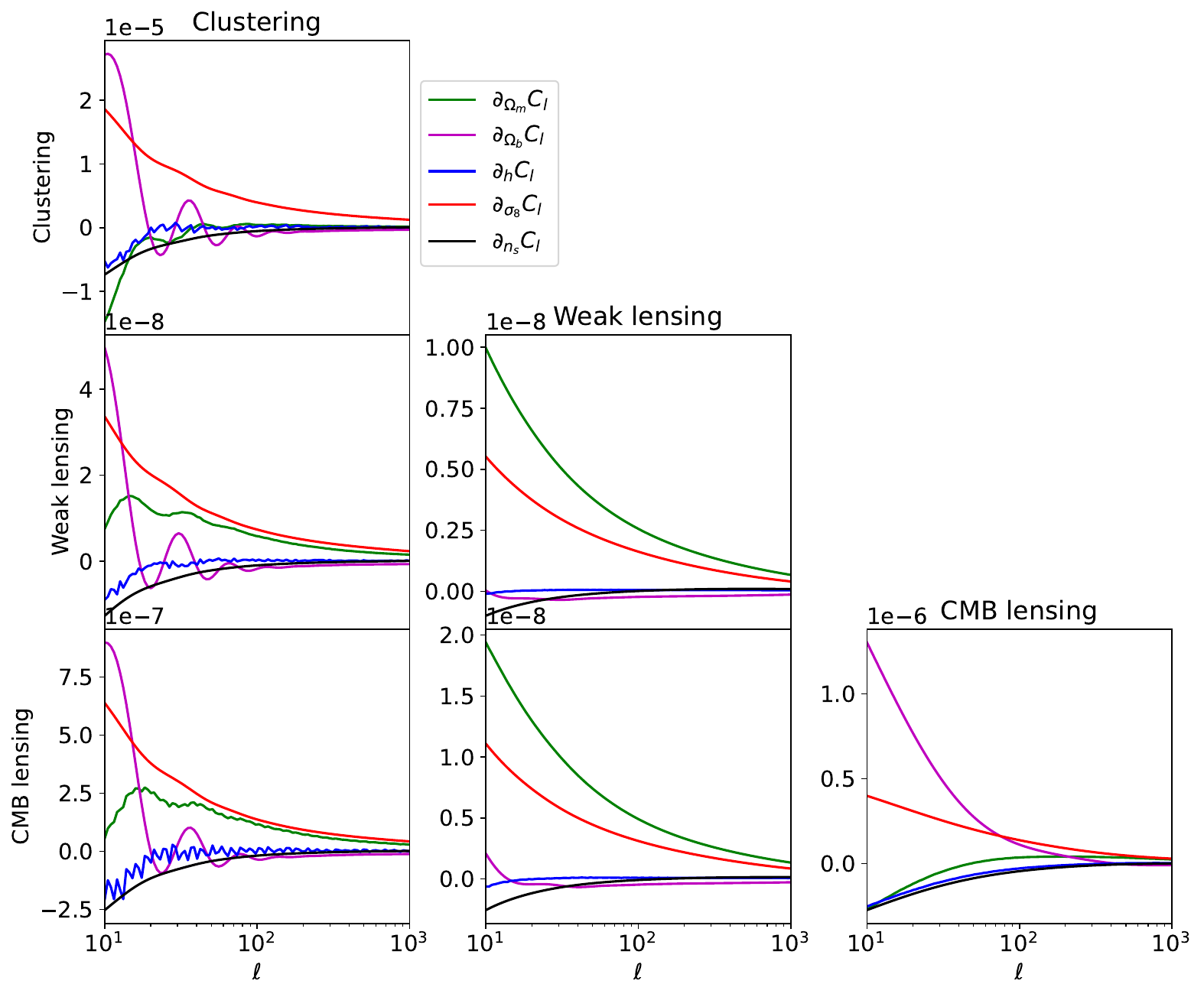} 
    \caption{Shows how the derivatives of auto- and cross-correlation angular power spectra of galaxy clustering, weak lensing and CMB lensing with respect the five \lcdm cosmological parameters $\Omega_{\rm{m}}$, $\Omega_{\rm{b}}$, $h$, $\sigma_{\rm{8}}$ and $n_{\rm{s}}$. The linear matter power spectrum was computed using the E\&H formula. In each panel, we over plot the five derivatives for a given angular power spectrum.}
   \label{fig:cls_diffs}
\end{figure*}

In Fig. \ref{fig:cls_diffs} we show the derivatives of auto- and cross-correlation angular power spectra of galaxy clustering, weak lensing and CMB lensing with respect the five \lcdm cosmological parameters. In each panel, we over plot the five derivatives for a given angular power spectrum. In all cases we employed the E\&H formula to compute the linear matter power spectrum. We can see that different cosmological parameters have vastly different impact in the computation of the angular power spectra. This can be seen not only in the different degrees of smoothness but also in the signs of the derivatives. We also note that the derivatives with respect to $h$ are jagged, specially when galaxy clustering is involved. This is caused by the numerical scheme used to integrate the redshift distribution.

\section{Tutorial} \label{App: Tutorial}

\begin{figure} 
    \includegraphics[width=\linewidth]{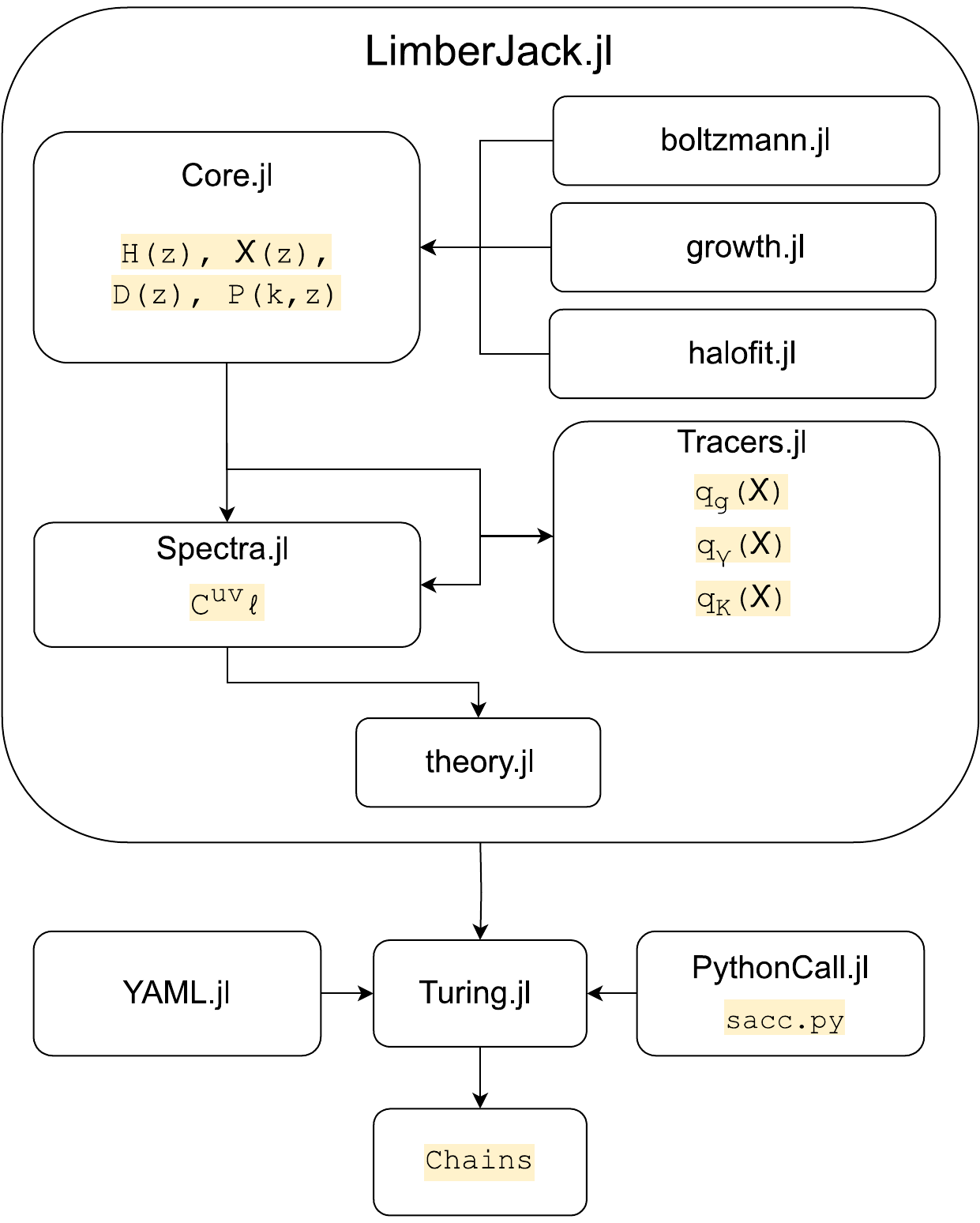} 
    \caption{Shows an schematic of \LB's code.}
   \label{fig:structure}
\end{figure}
In this section we will show a small demonstration of how to use \LB. We shall begin by having a quick overview of the structure of \LB. This will help us understand how to use the code later on. As described in Sect. \ref{Subsect: limeberJack}, \LB is a modular code. A summary of the relationship between the different modules can be found in Fig. \ref{fig:structure}. The core of \LB is the \texttt{Cosmology} structure and its homonymous constructor. \texttt{Cosmology}'s role is to contain all the information on how to compute the theoretical predictions. When the \texttt{Cosmology} structure is initiated it computes theoretical predictions for the expansion history, comoving distance, the growth factor and the matter power spectrum according to the provided prescriptions. The computation of background quantities occurs within \texttt{core.jl} itself. However, the computation of the matter power spectrum takes place across three different modules, \texttt{boltzmann.jl}, \texttt{growth.jl} and \texttt{halofit.jl}, which compute the primordial matter power spectrum, the linear growth factor and the non-linear corrections respectively. The different predictions are evaluated at a grid of values which are then used to build interpolators for each of the respective quantities. The interpolators are then stored inside the \texttt{Cosmology} structure. This allows the user to quickly compute the theoretical predictions at arbitrary values by using a series of public functions. These functions take the \texttt{Cosmology} structure and the necessary inputs. Inside these functions the corresponding interpolator is evaluated at the provided inputs to return the prediction to the user. 

Let us briefly showcase how this is done in code \ref{code:init} by computing some relatively straight forward predictions: 

\begin{lstlisting}[escapechar=|, language=julia,
label=code:init, caption= Shows basic use of LimberJack.jl]
# Install LimberJack
using Pkg
Pkg.add("LimberJack")

# Import it
using LimberJack

# create Cosmology instance
cosmo = Cosmology() 
# same as 
# cosmology = Cosmology(|$\Omega_{\rm{m}}=0.3$|, |$\Omega_b=0.05$|,
#                       h=0.67, ns=0.96, |$\sigma_{\rm{8}}=0.81$|)     

# Compute predictions
zs = [0.1, 0.5, 1.0, 3.0]
H = cosmo.cpar.h*100*Ez(cosmo, zs)
chi = comoving_radial_distance(cosmo, zs)
Dz = growth_factor(cosmo, zs)
fz = growth_rate(cosmo, zs)
fs8z = fs8(cosmo, zs)
\end{lstlisting} 

Computing the matter power spectrum is just as easy (see code \ref{code:Pk}). However, the computation is subject to a series of options that the user can alter. By default \texttt{Cosmology} will use the E\&H formula to find $P_0$ and it will not apply non-linear corrections.  These settings can be changed by specifying the keyword arguments \texttt{tk\_mode} and \texttt{Pk\_mode}. In terms of transfer functions \LB offers two possibilities  \texttt{tk\_mode} = \texttt{:EisHu} (default) / \texttt{:EmuPk} which correspond to using the Eisenstein and Hu formula or \texttt{EmuPk}. Similarly, \texttt{Pk\_mode} = \texttt{:linear} / \texttt{:Halofit} which determines whether or not non-linear corrections are applied using \texttt{Halofit}. \LB offers two distinct public functions to evaluate either the linear or non-linear matter power spectrum regardless of the choice in \texttt{Pk\_mode}. However, if \texttt{Pk\_mode} = \texttt{:linear} the two functions will return the linear matter power spectrum.

\begin{lstlisting}[escapechar=|, language=julia,
label=code:Pk, caption= Shows how to compute the matter power spectrum in LimberJack.jl]
using NPZ

cosmo_lin_EisHu = Cosmology(|$\Omega_{\rm{m}}=0.25$|, |$\Omega_b=0.03$|,
                             h=0.70, ns=1.0,
                             |$\sigma_{\rm{8}}=0.78$|,
                             tk_mode=:EisHu,
                             Pk_mode=:linear) 
                             
cosmo_nonlin_EisHu = Cosmology(|$\Omega_{\rm{m}}=0.25$|, |$\Omega_b=0.03$|,
                             h=0.70, ns=1.0,
                             |$\sigma_{\rm{8}}=0.78$|,
                             tk_mode=:EisHu,
                             Pk_mode=:Halofit) 

cosmo_nonlin_emupk = Cosmology(|$\Omega_{\rm{m}}=0.25$|, |$\Omega_b=0.03$|,
                             h=0.70, ns=1.0,
                             |$\sigma_{\rm{8}}=0.78$|,
                             tk_mode=:EmuPk,
                             Pk_mode=:Halofit)

zs = [0.1, 0.3, 0.5]
ks = [100, 300, 1000]
lin_eh_Pks = lin_Pk(cosmo_lin_EisHu, ks, zs) = nonlin_Pk(cosmo_lin_EisHu, ks, zs)
nonlin_eh_Pks = nonlin_Pk(cosmo_nonlin_EisHu, ks, zs)
nonlin_emupk_Pks = nonlin_Pk(cosmo_nonlin_emupk, ks, zs)
\end{lstlisting} 

Computing angular power spectra is a slightly more involved process. An example of how this is done in \LB can be found in code \ref{code:cls}. In this code we can see that first a \texttt{Cosmology} structure must be initiated. The \texttt{Cosmology} structure automatically computes the matter power spectrum given the user specifications. Then the user must compute the radial kernels of the relevant tracers by proving the corresponding \LB public functions with the \texttt{Cosmology} structure and the distribution of sources. Moreover, as described in \ref{Subsect: Angular Power Spectra} different tracers can be impacted by different systematic effects. These systematics are accounted by incorporating a series of nuisance parameters that can also be provided to the tracers public functions of \LB. The output of the tracer functions (see lines 10, 15 and 18 of code \ref{code:cls}) is a \texttt{Tracer} structure that hosts an interpolator for the corresponding radial kernel and the corrections described in Eqns. \ref{eq:ell_correction_1} and \ref{eq:ell_correction_2} (set to one in the case of clustering). The angular power spectra can computed by providing the \texttt{AngularCls} public function with the aforementioned \texttt{Cosmology} and \texttt{Tracer} objects as well as the desired multipoles.

\begin{lstlisting}[escapechar=|, language=julia,
label=code:cls, caption= Shows how to compute angular power spectra in LimberJack.jl]
# Initiate cosmology
cosmo = Cosmology()

# Define a distribution of sources
z = Vector(range(0., stop=2., length=256))
nz = @. exp(-0.5*((z-0.5)/0.05)^2)

# Create tracer objects
bias = 1.0 # Galaxy-mass bias
tg = NumberCountsTracer(cosmo, z, nz; b=bias)

mbias = 0.0 # shape multiplicative bias
A_IA = 0.0 # Amplitude of intrinsic alignments power spectrum
alpha_IA = 0.0 # Slope of intrinsic alignments power spectrum
ts = WeakLensingTracer(cosmo, z, nz;
                       m=mbias, IA_params=[A_IA, alpha_IA])

tk = CMBLensingTracer(cosmo)

# Compute power spectra
ls = [10.0, 30.0, 100.0, 300.0, 1000.0]
Cl_gg = angularCls(cosmo, tg, tg, ls)
Cl_gs = angularCls(cosmo, tg, ts, ls)
Cl_ss = angularCls(cosmo, ts, ts, ls)
Cl_gk = angularCls(cosmo, tg, tk, ls)
Cl_sk = angularCls(cosmo, ts, tk, ls)
\end{lstlisting}

Finally, in code \ref{code:turing} we show how to use \texttt{Turing.jl} in unison with \texttt{LimberJack} to build and sample an statistical model for a the DES Y1 3x2-pt analysis. The first step is to load the data. For this purpose we will use the libraries \texttt{YAML} and \texttt{sacc}, ubiquitous in astrophysics. \jl counts with a native implementation of \texttt{YAML} but not of \texttt{sacc}. However, calling \texttt{Python} libraries from \jl is extremely simple thanks to the \texttt{PythonCall.jl} library.  \texttt{PythonCall.jl} allows us to import \texttt{sacc.py} as a \jl module and read files entirely within \jl. Note that in order to this we must first install \texttt{sacc} in the \texttt{Python} environment of \jl or point \texttt{PythonCall} to our local \texttt{Pytthon} installation. Instructions on how to do this can be found in the \LB GitHub.  Once the data are loaded they must be passed to the \LB public function \texttt{make\_data} which turns the files into \jl structures that \LB can easily manage.

After that, the user can use \texttt{Turing.jl}'s $\texttt{\@model}$ macro to define an statistical model. Inside the model, the user must define the priors for the parameters of the model using the $\texttt{Distributions.jl}$ API. Note that while the cosmology parameters can be directly passed to the \texttt{Cosmology} structure constructor the nuisance parameters must be stored inside a \jl dictionary. The name of these parameters inside the dictionary must follow a strict convention "\texttt{tracer\_name} + \_\_ + \texttt{bin\_number}+ \_ +\texttt{nuisance\_parameter\_name}".

In order to obtain the theoretical prediction for the DES Y1 3x2-pt analysis data vector the user must provide the just initiated \texttt{Cosmology} structure and the \texttt{meta} and \texttt{files} structures generated by \texttt{make\_data} to the public function \texttt{Theory}. Moreover, the user must also provide the nuisance parameter dictionary using the \texttt{Nuisances} keyword argument. The \texttt{Theory} function orchestrates the computation of the theory vector, computing the necessary radial kernels and evaluating the angular power spectra in the correct order. 

Once a theoretical prediction has been obtained, the user must define how the data is distributed with respect to the theory prediction. In the case of a Gaussian likelihood, this corresponds to a multivariate Gaussian distribution with mean the theory prediction and covariance matrix the data covariance matrix. All is left is to do is to use \texttt{Turing.jl} to condition the model on the observations (see line 80 of code \ref{code:turing}), define the sampler we wish to use and sample the model. For a more thorough explanation of how to use \texttt{Turing.jl} and its different options please see \texttt{Turing.jl}'s documentation. The results can be found plotted in Fig.  \ref{fig:DESY1}

\begin{lstlisting}[escapechar=|, language=julia,
label=code:turing, caption= Shows how to perform cosmological inference by combining Turing.jl and LimberJack.jl]
# Imports
using LinearAlgebra
using Turing
using LimberJack
using YAML
using PythonCall
sacc = pyimport("sacc");

# Load data
sacc_path = "cls_FD_covG.fits"
yaml_path = "DESY1.yml"
sacc_file = sacc.Sacc().load_fits(sacc_path)
yaml_file = YAML.load_file(yaml_path)
meta, files = make_data(sacc_file, yaml_file)
data = meta.data
cov = meta.cov

# Define model
@model function model(data;
    meta=meta, 
    files=files)
    |$\Omega_{\rm{m}}$| ~ Uniform(0.2, 0.6)
    |$\Omega_b$| ~ Uniform(0.028, 0.065)
    h ~ TruncatedNormal(0.72, 0.05, 0.64, 0.82)
    |$\sigma_{\rm{8}}$| ~ Uniform(0.4, 1.2)
    ns ~ Uniform(0.84, 1.1)

    DESgc__0_b ~ Uniform(0.8, 3.0)
    DESgc__1_b ~ Uniform(0.8, 3.0)
    DESgc__2_b ~ Uniform(0.8, 3.0)
    DESgc__3_b ~ Uniform(0.8, 3.0)
    DESgc__4_b ~ Uniform(0.8, 3.0)
    DESgc__0_dz ~ TruncatedNormal(0.0, 0.007, -0.2, 0.2)
    DESgc__1_dz ~ TruncatedNormal(0.0, 0.007, -0.2, 0.2)
    DESgc__2_dz ~ TruncatedNormal(0.0, 0.006, -0.2, 0.2)
    DESgc__3_dz ~ TruncatedNormal(0.0, 0.01, -0.2, 0.2)
    DESgc__4_dz ~ TruncatedNormal(0.0, 0.01, -0.2, 0.2)
    DESwl__0_dz ~ TruncatedNormal(-0.001, 0.016, -0.2, 0.2)
    DESwl__1_dz ~ TruncatedNormal(-0.019, 0.013, -0.2, 0.2)
    DESwl__2_dz ~ TruncatedNormal(0.009, 0.011, -0.2, 0.2)
    DESwl__3_dz ~ TruncatedNormal(-0.018, 0.022, -0.2, 0.2)
    DESwl__0_m ~ Normal(0.012, 0.023)
    DESwl__1_m ~ Normal(0.012, 0.023)
    DESwl__2_m ~ Normal(0.012, 0.023)
    DESwl__3_m ~ Normal(0.012, 0.023)
    A_IA ~ Uniform(-5, 5) 
    alpha_IA ~ Uniform(-5, 5)

    nuisances = Dict("DESgc__0_b" => DESgc__0_b,
                     "DESgc__1_b" => DESgc__1_b,
                     "DESgc__2_b" => DESgc__2_b,
                     "DESgc__3_b" => DESgc__3_b,
                     "DESgc__4_b" => DESgc__4_b,
                     "DESgc__0_dz" => DESgc__0_dz,
                     "DESgc__1_dz" => DESgc__1_dz,
                     "DESgc__2_dz" => DESgc__2_dz,
                     "DESgc__3_dz" => DESgc__3_dz,
                     "DESgc__4_dz" => DESgc__4_dz,
                     "DESwl__0_dz" => DESwl__0_dz,
                     "DESwl__1_dz" => DESwl__1_dz,
                     "DESwl__2_dz" => DESwl__2_dz,
                     "DESwl__3_dz" => DESwl__3_dz,
                     "DESwl__0_m" => DESwl__0_m,
                     "DESwl__1_m" => DESwl__1_m,
                     "DESwl__2_m" => DESwl__2_m,
                     "DESwl__3_m" => DESwl__3_m,
                     "A_IA" => A_IA,
                     "alpha_IA" => alpha_IA,)

    cosmology = Cosmology(|$\Omega_{\rm{m}}$|=|$\Omega_{\rm{m}}$|, |$\Omega_b$|=|$\Omega_b$|,
                          h=h, ns=ns, |$\sigma_{\rm{8}}$|=|$\sigma_{\rm{8}}$|,
                          tk_mode=:EisHu,
                          Pk_mode=:Halofit)

    theory = Theory(cosmology, meta, files; Nuisances=nuisances)
    data ~ MvNormal(theory, cov)
end

# Condition model on data
cond_model = model(data)

# Define sampler
nadapts = 500
TAP = 0.65
sampler = NUTS(adaptation, TAP)

# Sample model 
iterations = 1000
chain = sample(cond_model, sampler, iterations)             
\end{lstlisting} 

\bibliographystyle{mnras}
\bibliography{main}

\end{document}